\documentclass[longauth]{aa} 

\usepackage{graphicx}
\usepackage{txfonts}
\usepackage{hyperref}
\hypersetup{
    colorlinks=true,
    linkcolor=blue,
    citecolor=blue,      
    urlcolor=cyan,
}
\usepackage{subcaption}

\begin{document} 

   \title{The GRAVITY young stellar object survey}
   \subtitle{V. The orbit of the T\,Tauri binary star WW\,Cha}

   \titlerunning{The orbit of the T\,Tauri binary star WW\,Cha}

   \author{  GRAVITY Collaboration\thanks{GRAVITY is developed in collaboration by the Max Planck Institute for Extraterrestrial Physics, LESIA of Paris Observatory, IPAG of Université Grenoble Alpes/CNRS, the Max Planck Institute for Astronomy, the University of Cologne, the Centro de Astrofísica e Gravitação, and the European Southern Observatory.}: F. Eupen
          \inst{1}
          \and
          L. Labadie
          \inst{1}
          \and
          R. Grellmann
          \inst{1}
          \and
          K. Perraut
          \inst{2}
          \and
          W. Brandner
          \inst{3}
          \and
          G. Duch\^ene
          \inst{4,2}
          \and
          R. K\"ohler
          \inst{5}
          \and
          J. Sanchez-Bermudez
          \inst{3,6}
          \and
          R. Garcia Lopez
          \inst{3,7,8}
          \and
          A. Caratti o Garatti
          \inst{3,7,8}
          \and
          M. Benisty
          \inst{2,9}
          \and
          C. Dougados
          \inst{2}
          \and
          P. Garcia
          \inst{10,11}
          \and
          L. Klarmann\inst{3}
          \and A. Amorim\inst{10, 16} 
          \and M. Baub\"ock\inst{13} 
          \and J.P. Berger\inst{2}  
          \and P. Caselli\inst{13} \and Y. Cl\'enet\inst{12} \and V. Coud\'e du Foresto\inst{12} \and P.T. de Zeeuw\inst{13, 17} \and A. Drescher\inst{13} 
          \and G. Duvert\inst{2} \and A. Eckart\inst{1, 15} \and F. Eisenhauer\inst{13}  \and M. Filho\inst{10,11} \and V. Ganci\inst{1} \and F. Gao\inst{13} \and E. Gendron\inst{12} \and R. Genzel\inst{13} \and S. Gillessen\inst{13} \and G. Heissel\inst{12} 
          \and Th. Henning\inst{3}
          \and S. Hippler\inst{3}
          \and M. Horrobin\inst{1} \and Z. Hubert\inst{2} \and A. Jim\'enez-Rosales\inst{13} \and L. Jocou\inst{2} \and P. Kervella\inst{12} \and S. Lacour\inst{12} 
          \and V. Lapeyr\`ere\inst{12} 
          \and J.B. Le Bouquin\inst{2}
          \and P. L\'ena\inst{12} \and T. Ott\inst{13} \and T. Paumard\inst{12} \and G. Perrin\inst{12} \and O. Pfuhl\inst{14} 
          \and G. Rodr\'iguez-Coira\inst{12} \and G. Rousset\inst{12}
          \and S. Scheithauer\inst{3}
          \and J. Shangguan\inst{13} \and T. Shimizu\inst{13} \and J. Stadler\inst{13} \and O. Straub\inst{13} \and C. Straubmeier\inst{2} 
          \and E. Sturm\inst{13} \and E. van Dishoeck\inst{13, 17} \and F. Vincent\inst{12} \and S.D. von Fellenberg\inst{13} \and F. Widmann\inst{13} \and J. Woillez\inst{14} \and A. Wojtczak\inst{1}
    }

\institute{
I. Physikalisches Institut, Universit{\"a}t zu K{\"o}ln, Z{\"u}lpicher Str. 77, 50937 K{\"o}ln, Germany\\
\email{eupen@ph1.uni-koeln.de}
\and Univ. Grenoble Alpes, CNRS, IPAG, 38000 Grenoble, France
\and Max Planck Institute for Astronomy, K\"onigstuhl 17, 69117 Heidelberg, Germany
\and Astronomy Department, University of California, Berkeley, CA 94720-3411, USA
\and
University of Vienna, Department of Astrophysics, T\"urkenschanzstr. 17, A-1180 Vienna, Austria
\and
Instituto de Astronom\'ia, Universidad Nacional Aut\'onoma de M\'exico, Apdo. Postal 70264, Ciudad de M\'exico, 04510, M\'exico
\and
Dublin Institute for Advanced Studies, 31 Fitzwilliam Place, D02\,XF86 Dublin, Ireland
\and
School of Physics, University College Dublin, Belfield, Dublin 4, Ireland\and
Unidad Mixta Internacional Franco-Chilena de Astronom\'ia (CNRS UMI 3386), Departamento de Astronom\'ia, Universidad de Chile,
Camino El Observatorio 1515, Las Condes, Santiago, Chile
\and
CENTRA, Centro de Astrof\'isica e Gravita\c c\~{a}o, Instituto Superior T\'ecnico, Avenida Rovisco Pais 1, 1049 Lisboa, Portugal
\and
Universidade do Porto, Faculdade de Engenharia, Rua Dr. Roberto Frias, 4200-465 Porto, Portugal
\and
LESIA, Observatoire de Paris, PSL Research University, CNRS, Sorbonne Universit\'es, UPMC Univ. Paris 06, Univ. Paris Diderot,Sorbonne Paris Cit\'e, France
\and
Max Planck Institute for Extraterrestrial Physics, Giessenbachstrasse, 85741 Garching bei M\"unchen, Germany
\and
European Southern Observatory, Karl-Schwarzschild-Str. 2, 85748 Garching, Germany
\and
Max-Planck-Institute for Radio Astronomy, Auf dem H\"ugel 69, 53121 Bonn, Germany
\and
Universidade de Lisboa - Faculdade de Ci\^{e}ncias, Campo Grande, 1749-016 Lisboa, Portugal
\and
Sterrewacht Leiden, Leiden University, Postbus 9513, 2300 RA Leiden, The Netherlands
}

   \date{Received ; accepted}

 
  \abstract
   {Close young binary stars are unique laboratories for the direct measurement of pre-main-sequence (PMS) stellar masses and their comparison to evolutionary theoretical models. At the same time, a precise knowledge of their orbital parameters when still in the PMS phase offers an excellent opportunity for understanding the influence of dynamical effects on the morphology and lifetime of the circumstellar as well as circumbinary material.}
   {The young T\,Tauri star WW\,Cha was recently proposed to be a close binary object with strong infrared and submillimeter excess associated with circum-system emission, which makes it dynamically a very interesting source in the above context. The goal of this work is to determine the astrometric orbit and the stellar properties of WW\,Cha using multi-epoch interferometric observations.}
   {We derive the relative astrometric positions and flux ratios of the stellar companion in WW\,Cha from the interferometric model fitting of observations made with the VLTI instruments AMBER, PIONIER, and GRAVITY in the near-infrared from 2011 to 2020. For two epochs, the resulting uv-coverage in spatial frequencies permits us to perform the first image reconstruction of the system in the K band. The positions of nine epochs are used to determine the orbital elements and the total mass of the system. Combining the orbital solution with distance measurements from Gaia DR2 and the analysis of evolutionary tracks, we constrain the mass ratio.}
   {We find the secondary star orbiting the primary with a period of $T$\, =\,206.55~days, a semimajor axis of $a$\,=\,1.01~au, and a relatively high eccentricity of $e$\,=\,0.45. The dynamical mass of $M_{\mathrm{tot}}$\,=\,3.20$\,M_{\odot}$ can be explained by a mass ratio between $\sim$\,0.5 and 1, indicating an intermediate-mass T~Tauri classification for both components. 
   The orbital angular momentum vector is in close alignment with the angular momentum vector of the outer disk as measured by ALMA and SPHERE, resulting in a small mutual disk inclination. 
   The analysis of the relative photometry suggests the presence of infrared excess surviving in the system and likely originating from truncated circumstellar disks. 
   The flux ratio between the two components appears variable, in particular in the K band, and may hint at periods of triggered higher and lower accretion or changes in the disks' structures.}
   {
   The knowledge of the orbital parameters, combined with a relatively short period, makes WW\,Cha an ideal target for studying the interaction of a close young T\,Tauri binary with its surrounding material, such as time-dependent accretion phenomena. Finding WW\,Cha to be composed of two (probably similar) stars led us to reevaluate the mass of WW\,Cha, which had been previously derived under the assumption of a single star. This work illustrates the potential of long baseline interferometry to precisely characterize close young binary stars separated by a few astronomical units. Finally, when combined with radial velocity measurements, individual stellar masses can be derived and used to calibrate theoretical PMS models.
   }


   \maketitle
%
\section{Introduction}
\label{sec:intro}
While the formation of T\,Tauri stars is generally well understood, masses and ages of pre-main-sequence (PMS) stars are usually derived via the comparison of their positions in the Hertzsprung-Russell diagram (HRD) with theoretical models of PMS stellar evolution \citep[e.g., ][]{2000A&A...358..593S,Dotter2008,Tognelli2011,2015A&A...577A..42B}.  However, despite the improved accuracy of such theoretical models, the early stages of PMS stellar evolution still remain relatively uncertain.
Furthermore, the derivation of stellar masses from the evolutionary tracks requires accurate photometry and extinction estimates. 
Model-independent dynamical masses can be measured in binary systems, which requires the determination of the orbital parameters. Complemented with radial velocity (RV) spectroscopic measurements and knowledge of the orbit inclination, the mass of the individual components can be obtained. 
While at least 50\% of solar mass type stars are found to be in binary systems \citep[][and references therein]{2010ApJS..190....1R, Duchene2013}, the set of orbital parameters is only known, from multi-epoch monitoring, for a limited number of T Tauri binary stars  (e.g., \citet{Mathieu2007b,Schaefer2014, Kellogg2017}, and references therein). Because of the potentially long orbital timescales, close binary stars with periods shorter than a few years are the preferred targets for orbit determination. These are, for instance, spectroscopic binaries with periods on the order of tens of days that are characterized in spectro-photometry campaigns (e.g., UZ\,Tau\,E and DQ\,Tau; \citet{Martin2005,2016ApJ...818..156C}) and binaries with a few au separation (e.g., GG\,Tau\,Ab; \citet{DiFolco2014}) characterized through high-angular-resolution techniques.

Another interesting aspect concerning close young binary systems regards the study of the interaction between the orbiting pair and the long-lived circum-system environment, which may be dynamically influenced by the binary star. While for the widest binaries (a$\gtrsim$100\,au) each component may host a circumstellar disk with properties similar to those of the disks around single stars \citep{Mathieu2007a}, tidal forces and truncation effects occur in the existing circum-stellar and -binary disks around binaries with a few tens of au separation, leading to peculiar morphologies and/or configurations \citep{Miranda2015,Kurtovic2018,Kennedy2019,Czekala2019}, or possibly to reduced disk lifetimes \citep{Cieza2009}. In this context, our comprehension of disk properties and evolution in close binaries raises important questions regarding the as-of-yet poorly understood processes of planet formation in close binary systems \citep{Hatzes2003ApJ,Chauvin2011,Doyle2011,Welsh2012}. The influence of close binarity on key stellar processes -- resulting, for instance, in pulsed gas accretion \citep{Tofflemire2017,2019ApJ...877...29M} -- is another example of interaction involving disk evolution and multiplicity. In both cases, the clear knowledge of the binary orbital solution is a strong prior for a correct interpretation of the observations.

In this work, we study \object{WW Cha}; it is a young variable PMS star located in the Chamaeleon I molecular cloud and is one of the brightest nearby T Tauri stars.
The star is located at the revised distance of $191.0^{+1.3}_{-1.3}$~pc (\citet{2018AJ....156...58B}, cf. Appendix~\ref{sec:parallax}), slightly farther than the previously assumed distance of $\sim$160\,pc for Chamaeleon\,I \citep{1997A&A...327.1194W}, and has a median age of $\sim$1\,Myr (\citet{2007ApJS..173..104L} determined after revising the distance of the cloud to $\sim$190\,pc; see, e.g., \citet{2010arXiv1006.3676K} and \citet{2011A&A...527A.145B}). Spectroscopic measurements suggest a spectral type between K5 \citep[$T_\ast$=4350\,K;][]{2007ApJS..173..104L} and K0 \citep[$T_\ast$=5110\,K;][]{2016A&A...585A.136M}.
\citet{Manoj2011} report a strong infrared excess for that source, which is attributed to circumstellar emission from the hosted protoplanetary disk. 
Thanks to Very Large Telescope Interferometer (VLTI) long baseline interferometry, WW\,Cha was found to be a binary star by \citet{2015A&A...574A..41A}, who measured a flux ratio of $\sim$0.6 in the H band and an angular separation between 4.7 and 6.4 milliarcseconds (mas) in three epochs. WW~Cha cannot therefore be spatially resolved with current single-dish telescopes. 

We have acquired new interferometric observations of WW\,Cha with the instruments AMBER \citep{Petrov2007} and GRAVITY \citep{Eisenhauer2017} and combine them with existing PIONIER \citep{LeBouquin2011} archival observations. With this unique multi-epoch high-resolution dataset, we conduct a detailed astrometric study of the binary star to determine the orbital solution for the pair. 
Section~\ref{sec:observations} describes the observations and data reduction. In Sect.~\ref{sec:results_modeling}, we present the results of our modeling of the binary astrometry and photometry, as well as report the derived orbital solution. We discuss our results and perspectives in Sect.~\ref{sec:results} and finally conclude the paper in Sect.~\ref{sec:conclusion}.
%
\section{Observations and data reduction}
\label{sec:observations}
The data were taken in the H and K bands between 2011 and 2020 with the VLTI, which can combine the light of up to four telescopes using either the 8.2-m Unit Telescopes (UTs) or the relocatable 1.8-m Auxiliary Telescopes (ATs). The observable quantities measured with the interferometer are the (squared) visibilities, which inform us on the object's characteristic size, and the closure phases, which indicate how the intensity distribution of the object departs from a centrosymmetric geometry \citep{Monnier2003}. 
\vspace{0.2cm}\\
{\it PIONIER. }
Observations of WW\,Cha were obtained with four ATs in 2011, 2012, and 2015. The 2011 and 2012\footnote{The 2011 and 2012 data are not available via the ESO data archive. The calibrated data (as published in \citet{2015A&A...574A..41A}) were downloaded from the PIONIER consortium website.} data were acquired in undispersed mode in the H band, whereas the 2015 data were in part acquired in dispersed mode with a spectral resolution of $R$\,$\sim$\,30 at 1.65\,$\mu$m. The six baselines span a range of 22\,m to 123\,m, which corresponds to a maximal spatial resolution $\lambda/2B$ of $\sim$\,1.4\,mas ($\sim$\,0.3\,au) at the distance of the source (i.e., 191\,pc). The data were reduced and calibrated using the PIONIER data reduction software PNDRS \citep{LeBouquin2011}.
\vspace{0.2cm}\\
{\it AMBER. }
WW\,Cha was observed with AMBER and three UTs in 2009 (082.C-0920(A)) and 2017 (098.C-0334(F,G)) in low-resolution mode ($R\sim$35) across the H and K bands simultaneously. However, due to a technical failure of one of the telescopes in 2009 (hence offering only one baseline in that case) and bad weather conditions for the first run in 2017, we only retained the second AMBER run from 2017 for this study. Furthermore, difficulties in the absolute calibration of the data in the H band led us to only exploit the low-resolution K band signal here. The UT2-UT3-UT4 triplet offers a spatial resolution of 2.6\,mas ($\sim$\,0.5\,au) in the K band. To reduce and calibrate the data, the software package {\ttfamily amdlib} \citep{2007A&A...464...29T,2009A&A...502..705C} was used. We selected 20\% of the frames with the highest signal-to-noise ratio (S/N) for both calibrator and science data. We then calibrated the science measurement with the transfer function derived from two observations of the associated unresolved calibrator.
 \vspace{0.2cm}\\
{\it GRAVITY. }
WW\,Cha was observed with GRAVITY and the UTs in 2019 (one epoch), delivering a maximum spatial resolution of $\sim$2.0\,mas ($\sim$0.4\,au). The source was further observed in 2020 for two different epochs using the AT medium configuration that delivers a resolution of $\sim$2.5\,mas ($\sim$0.5\,au).
The data were reduced using the GRAVITY data reduction pipeline \citep{2014SPIE.9146E..2DL}. 

All GRAVITY data were acquired simultaneously at high spectral resolution ($R$\,$\sim$\,4000, 1742 channels across the K band from 2.0 to 2.4\,$\mu$m) in the science instrument and at low spectral resolution (five channels across the K band) with the fringe tracker (FT) operating at 1.2\,kHz (0.3\,kHz for parts of the 2020 observations)
to minimize the impact of the atmospheric turbulence. 
For each epoch, we checked the outcome of the data reduction by comparing the squared visibilities and closure phases measured with the science and FT instruments. We find a high consistency between the two datasets.

In this study, we focus our efforts on modeling the continuum emission, which corresponds to the thermal flux of the central stellar pair and its circumstellar environment. Within the data reduction pipeline, the high-resolution spectral data were re-binned from 1742 to 22 channels to increase the S/N over the continuum. This provides low-resolution spectral data that are more homogeneous across the band, with smaller error bars, and which are cleaned of data reduction outliers. In Appendix~\ref{sec:binned}, we show the comparison between the original and binned datasets for a subset of our GRAVITY uv-points. Following \cite{Perraut2019}, we considered, in some cases, a floor value for the error bars, computed by the pipeline, of $\pm$0.02 and $\pm$1$^{\circ}$ on the squared visibilities and on the closure phases, respectively. This was applied when the error bars computed by the pipeline appeared to be unreasonably small in order to avoid the effect of underestimated or correlated uncertainties. 
The stability of the GRAVITY transfer function and the overall accuracy on the interferometric observables allowed us to process the two datasets from January 2020 as two separate epochs for the orbit determination despite them being obtained within a short time span (see Sect.~\ref{sec:positions}). \\
\\
The observation log can be found in Table~\ref{tab:obs_log} together with the reduced data in Figs.~\ref{fig:data1} and \ref{fig:data2}. The logs describe the sequence of observation of the scientific target with the corresponding calibrators, the instrument, and the associated array configuration. Key information on the weather conditions is also provided. The figures show the measured squared visibilities, closure phases, and uv-coverage for all the datasets.\\
%
\section{Modeling and results}\label{sec:results_modeling}
\subsection{Astrometry and relative photometry}\label{sec:positions}

\begin{table*}[t!]
        \caption{Astrometry and photometry of the WW\,Cha binary system as determined from interferometric observations.}
        \centering
        \begin{tabular}{cccccccccc}
                \hline \hline
                  Date & Run      &Instrument & $\alpha$[mas] & $\beta$[mas] & $f_{\mathrm{comp}}/f_{\mathrm{prim}}$ & $f_{\mathrm{ext}}/f_{\mathrm{prim}}$ & $\chi^2_{min}$($\nu$) & $m_{H,prim}$ & $m_{H,comp}$  \\ \hline
                 10/2/11 & A & PIONIER     &  -2.48$^{+0.05}_{-0.04}$   &   5.12$^{+0.07}_{-0.07}$   & 0.61$^{+0.02}_{-0.02}$  & 0.23$^{+0.05}_{-0.05}$ &   2.48(6) & 7.87$^{+0.12}_{-0.09}$ & 8.41$^{+0.14}_{-0.14}$ \vspace{0.15cm}\\
                 6/3/12 & B & PIONIER      &  -4.84$^{+0.14}_{-0.12}$   &   4.05$^{+0.06}_{-0.07}$   & 0.69$^{+0.06}_{-0.04}$  & 0.46$^{+0.14}_{-0.13}$ &   2.06(6) & 8.04$^{+0.25}_{-0.25}$ & 8.44$^{+0.4}_{-0.4}$ \vspace{0.15cm}\\
                 2/7/12 & C & PIONIER               &  -2.99$^{+0.05}_{-0.05}$   &  -3.58$^{+0.07}_{-0.07}$   & 0.68$^{+0.03}_{-0.03}$  & 0.29$^{+0.04}_{-0.04}$ &   0.93(6) & 7.95$^{+0.11}_{-0.11}$ & 8.36$^{+0.18}_{-0.18}$  \vspace{0.15cm}\\
                 5/2/15 & D & PIONIER               &  -1.43$^{+0.03}_{-0.03}$   &   5.28$^{+0.03}_{-0.03}$   & 0.70$^{+0.02}_{-0.02}$  & 0.45$^{+0.01}_{-0.01}$ &   1.68(61) & 8.04$^{+0.09}_{-0.09}$ & 8.43$^{+0.12}_{-0.12}$  \vspace{0.15cm}\\
                 13/2/15 & E & PIONIER              &  -0.55$^{+0.04}_{-0.04}$   &   5.37$^{+0.05}_{-0.05}$   & 0.59$^{+0.04}_{-0.04}$  & 0.51$^{+0.03}_{-0.03}$ &   1.26(19) & 8.02$^{+0.10}_{-0.10}$ & 8.59$^{+0.22}_{-0.22}$  \vspace{0.15cm}\\
                \hline\hline
                   & &             &                            &                            &                         &                         &        & $m_{K,prim}$ & $m_{K,comp}$ \\
                \hline
                 9/3/17 & F & AMBER       &  -5.99$^{+0.03}_{-0.05}$            &   1.13$^{+0.26}_{-0.25}$   & 0.61$^{+0.01}_{-0.01}$   & 0.33$^{+0.02}_{-0.02}$ &   0.37(56) &  6.80$^{+0.06}_{-0.06}$ & 7.34$^{+0.07}_{-0.07}$ \vspace{0.15cm}\\
                 20/6/19 & G & GRAVITY    &  -5.777$^{+0.002}_{-0.002}$         &   1.97$^{+0.01}_{-0.01}$   & 0.505$^{+0.001}_{-0.001}$& 0.423$^{+0.007}_{-0.007}$& 8.89(1316) &  6.79$^{+0.05}_{-0.05}$ & 7.53$^{+0.05}_{-0.05}$ \vspace{0.15cm}\\
                & & (squeeze)             &  -5.790$^{+0.045}_{-0.045}$         &   1.94$^{+0.04}_{-0.04}$   & 0.611$^{+0.041}_{-0.041}$ & - &   1.06(1316) &  - & - \vspace{0.15cm}\\
                 28/1/20 & H & GRAVITY    &  -5.040$^{+0.008}_{-0.008}$         & 3.325$^{+0.01}_{-0.01}$    & 0.857$^{+0.004}_{-0.004}$& 0.42$^{+0.02}_{-0.02}$ &   9.17(185) &  6.97$^{+0.05}_{-0.05}$ & 7.14$^{+0.05}_{-0.05}$ \vspace{0.15cm}\\
                 30/1/20 & I  & GRAVITY   &  -4.875$^{+0.002}_{-0.002}$         &   3.473$^{+0.002}_{-0.002}$& 0.813$^{+0.002}_{-0.002}$& 0.360$^{+0.005}_{-0.005}$&12.17(1336) &  6.92$^{+0.05}_{-0.05}$ & 7.15$^{+0.05}_{-0.05}$ \vspace{0.15cm}\\
                & & (squeeze)             &  -4.854$^{+0.029}_{-0.029}$         &   3.473$^{+0.032}_{-0.032}$& 0.833$^{+0.040}_{-0.040}$ & - &  7.12(1336) &  - & - \vspace{0.15cm}\\
                \hline
        \end{tabular}
        \tablefoot{Best-fit results of the astrometry and relative photometry of WW\,Cha with 1$\sigma$ errors.
        For three PIONIER epochs (runs A, B, and E), multiple minima are found (cf. Sect.~\ref{sec:orbit}).
 For the three GRAVITY points, the systematic astrometric errors are estimated for $\alpha$ to be 0.02, 0.01, and 0.01\,mas, and for $\beta$ to be 0.02, 0.05, and 0.01\,mas, respectively (see text). 
 For the two GRAVITY epochs with independent image reconstruction (runs G and I), the relative positions and flux ratios resulting from the Gaussian fits (cf. Sect.~\ref{sect:imagerecon}) are shown as well.
The $\chi^2_{min}$ is the reduced $\chi^2$-value of the best fit, taking the degrees of freedom $\nu$ into account.
The last two columns indicate the H and K band apparent magnitudes for the primary and secondary, assuming the unchanged 2\,MASS total magnitudes of $m_H$=7.21$\pm$0.08 and $m_K$=6.08$\pm$0.05. In most cases, the relative magnitudes are limited by the 2\,MASS uncertainty. The technical details concerning the runs are given in the appendix. 
        }
        \label{tab:position_fits}
\end{table*}

In order to fit the binary position for each available epoch, we followed the approach in \citet{2015A&A...574A..41A}, using a model consisting of two point sources and an extended source in the form of a fully resolved component. While the point sources mostly represent the two stars, part of the flux attributed by the model to the point sources very likely comes from circumstellar material as well. The extended emission component is meant to represent the flux coming from all other sources that do not directly surround the two stellar components. This flux can be attributed to scattered light, flux coming from a circumbinary disk, or emission from connecting streamers. 
Adding all these options individually would unnecessarily complicate the model when the goal is to find the relative binary positions; they are thus combined into one smooth component.

To find the best fit of this model to the measured squared visibilities and closure phases for each epoch, we first conducted an extensive grid search over the following four parameters: the two relative positions of the secondary ($\alpha$, $\beta$), the flux ratio (relative to the primary) of the secondary, $f_2$=$F_{\mathrm{comp}}/F_{\mathrm{prim}}$, and the flux ratio of the extended source, $f_3$=$F_{\mathrm{ext}}/F_{\mathrm{prim}}$. For $\alpha$ and $\beta,$ we cover a range from $-15$ to $+15$\,mas, with a step size of $0.5$\,mas. The flux ratios $f_2$ and $f_3$ were scanned in step sizes of 0.1 from 0 to 1. From this four-dimensional grid, we produced, for each epoch, a two-dimensional $\chi^2$-map for the relative position of the secondary, with each pixel representing the best $\chi^2$-value with respect to the flux ratios at the given position (see Figs.~\ref{fig:chi2map_A} to \ref{fig:chi2map_I}). After the best parameters in the grid were identified, we performed a Markov chain Monte Carlo \cite[MCMC,][]{2013PASP..125..306F} simulation for each epoch, with the initial walkers distributed closely around the previously identified parameter values. In this way, we determined the highest likelihood for the four parameters with high precision and derived the error estimate at the 1$\sigma$ confidence interval (see Fig.~\ref{fig:positions_MCMC}). As one can see from Fig.~\ref{fig:chi2maps_grid}, the epochs A, B, and E feature several local $\chi^2$-minima with comparable values. Determining the true position requires additional prior knowledge, as presented in Sect.~\ref{sec:orbit}. 

With this approach, we confirm the results of \citet{2015A&A...574A..41A}, finding very similar relative positions for epochs A, B, and C, and further determine the position and flux ratios of the companion for six additional epochs, which indeed can only be explained by a gravitationally bound binary star system. The results can be found in Table~\ref{tab:position_fits} and are discussed in Sect.~\ref{sec:results}.

The uncertainties reported in our table are the 1$\sigma$ statistical error bars resulting from the fit presented in Fig.~\ref{fig:chi2maps_grid} and implemented for the described geometrical model. In some cases, and in particular for the GRAVITY points, the uncertainty is less than $\sim$\,10\,$\mu$as. We decided to investigate a more conservative estimate of the uncertainties on the position of the secondary, which is ultimately used for the derivation of the orbital parameters. For this purpose, we tested our data fit with slightly modified geometrical models in which, for instance, we added ring structures or replaced the point sources with Gaussian functions. The goal was to constrain the adopted astrometric accuracy by better accounting for the simplicity of our geometrical source model. 
After discarding solutions with a $\chi^2$-value that differs by more than twice the best fit, we estimated, for each epoch, the systematic astrometric error as the median of the deviations from the best fit shown in Table~\ref{tab:position_fits}. 
For AMBER and PIONIER, we find that the systematic error is on the same order of, or smaller than, the statistical error. For GRAVITY, the systematic error is found to range between 10 and 50\,$\mu$as for the six $\alpha$, $\beta$ positions. These values are adopted as uncertainties for the fit of the orbital parameters in Sect.~\ref{sec:orbit}.

\subsection{Determination of the orbital solution}
\label{sec:orbit}
\begin{figure}
        \includegraphics[width=0.92\columnwidth]{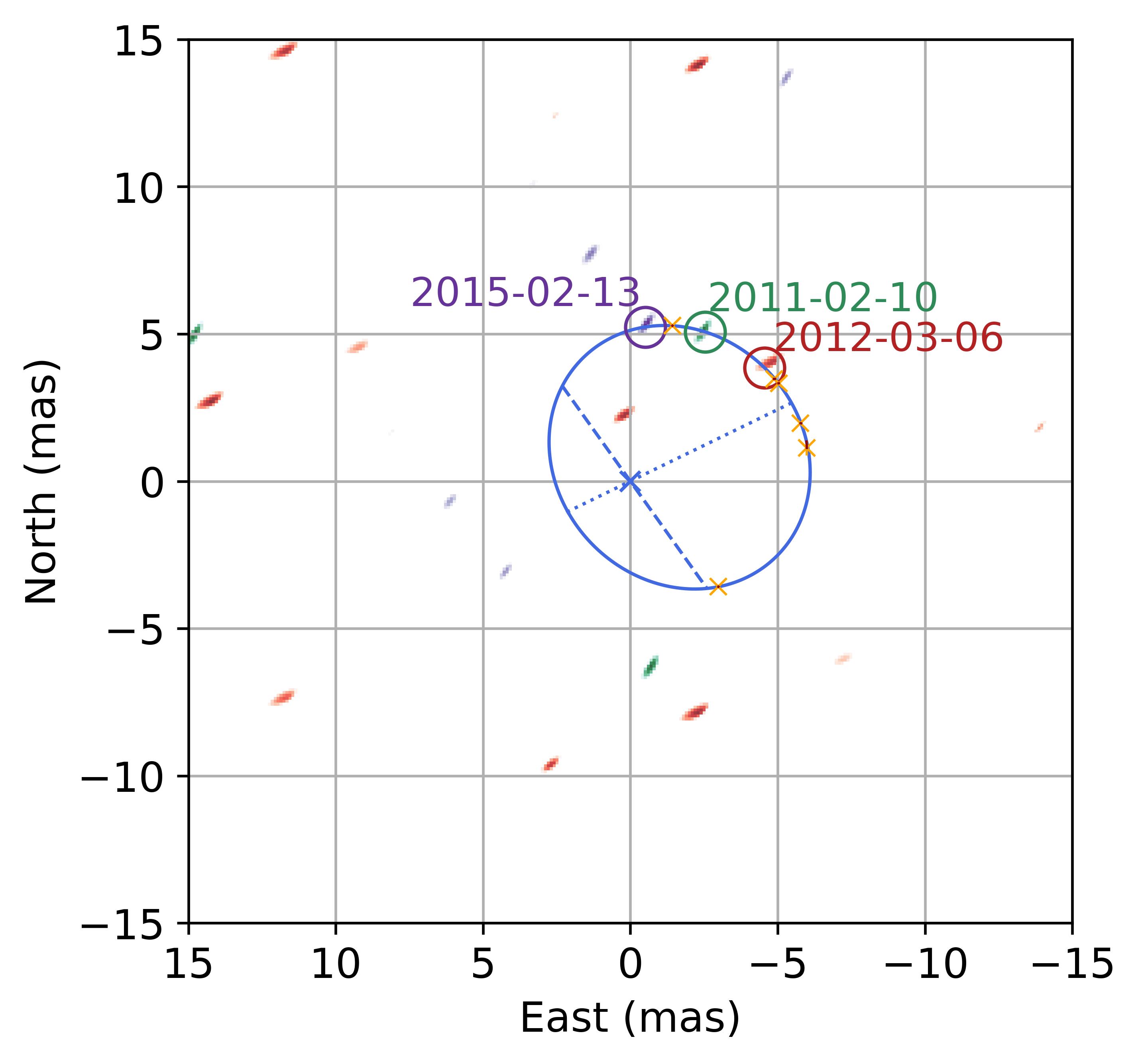}
    \caption{Best-fit orbit (blue; rotation counterclockwise) of the six unambiguous epochs (C, D, F, G, H, and I), marked by yellow crosses and error bars. The error is the $1\sigma$ confidence interval of the positions given in Table~\ref{tab:position_fits}. Most error bars are smaller than the representing crosses. The dashed line represents the line of nodes, and the dotted line connects peri- and apoastron. 
    The $\chi^2$-maps of the positions of the secondary for epochs A, B, and E are shown in green, red, and purple, respectively. Encircled are the local minima that visually match the derived orbit the best, with the center of each circle at the expected position for each epoch. 
    }
    \label{fig:chi2_orbit_abd13}
\end{figure}

As shown in the previous section, the relative positions for six of the nine available epochs can be determined unambiguously. Thus, we first determined the most likely orbital solution using only these six unambiguous identified positions. To do so, we scanned over a large range of orbital parameters, applying both a grid search \citep[as, e.g., in][]{Koehler2013} and an MCMC-based approach. Both methods were followed by a local optimization \citep[BFGS algorithm,][]{NoceWrig06, 2020SciPy-NMeth} and resulted in the same orbital parameters, which were applied to draw the intermediate orbit solution shown in Fig.~\ref{fig:chi2_orbit_abd13}.

In the next step, we first identified the possible positions of the companion that best match the intermediate orbit in the $\chi^2$-maps of runs A, B, and E. 
All three epochs show multiple, but very localized, $\chi^2$-minima with comparable values. 
Figure~\ref{fig:chi2_orbit_abd13} compares the expected positions -- according to the six-epoch intermediate orbit -- to the $\chi^2$-maps of the position fits for the three ambiguous epochs, A, B, and E. 
As can be seen for each of these three epochs, the predicted position clearly coincides with one of the local minima in the $\chi^2$-map, which is thus assumed to be the correct minimum and reported in Table~\ref{tab:position_fits}. The ratio\footnote{The subscript "alt" refers to  "alternative."} $\chi^2_{best}/\chi^2_{alt}$ (after local optimization) of the best matching position to the next best local $\chi^2$-minimum is $1.00$ for epoch A, $1.33$ for epoch B, and $0.69$ for epoch E. A value larger than one indicates that the chosen position is not the one that fits the data best. This is only the case for epoch B; however, for this and the other two epochs, none of the other possible positions are close enough to the intermediate orbit solution to be considered.

\begin{figure}
        \includegraphics[width=0.92\columnwidth]{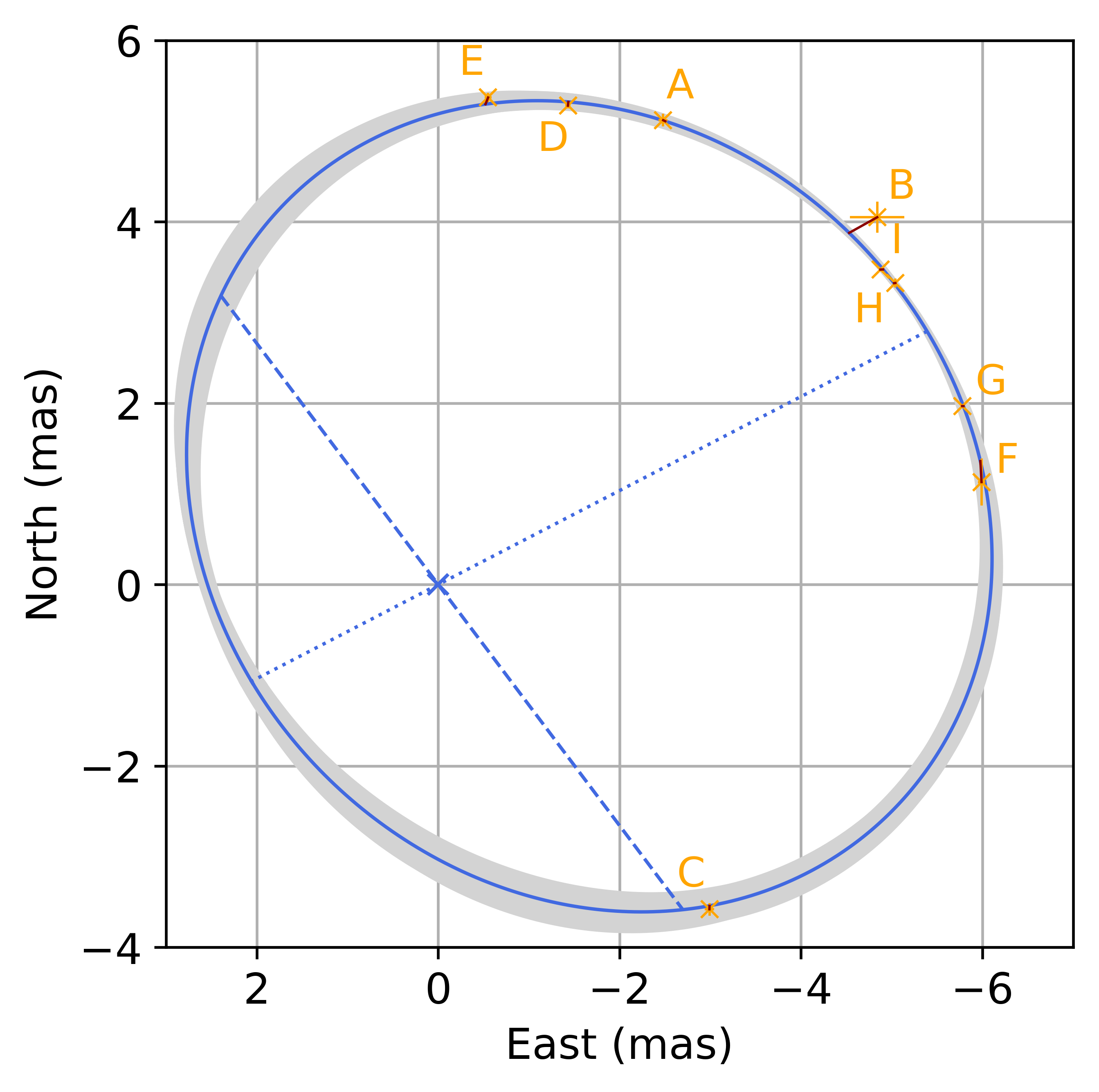}
    \caption{MCMC result of the orbit (blue) using all nine epochs of Table~\ref{tab:position_fits}, with a $1\sigma$ error. The red segments illustrate the residual of the fit. The gray area represents the $3\sigma$ confidence region. The rest is the same as in Fig. \ref{fig:chi2_orbit_abd13}.}
    \label{fig:orbit_all9_e02}
\end{figure}

\begin{figure*}
        \includegraphics[width=0.90\textwidth]{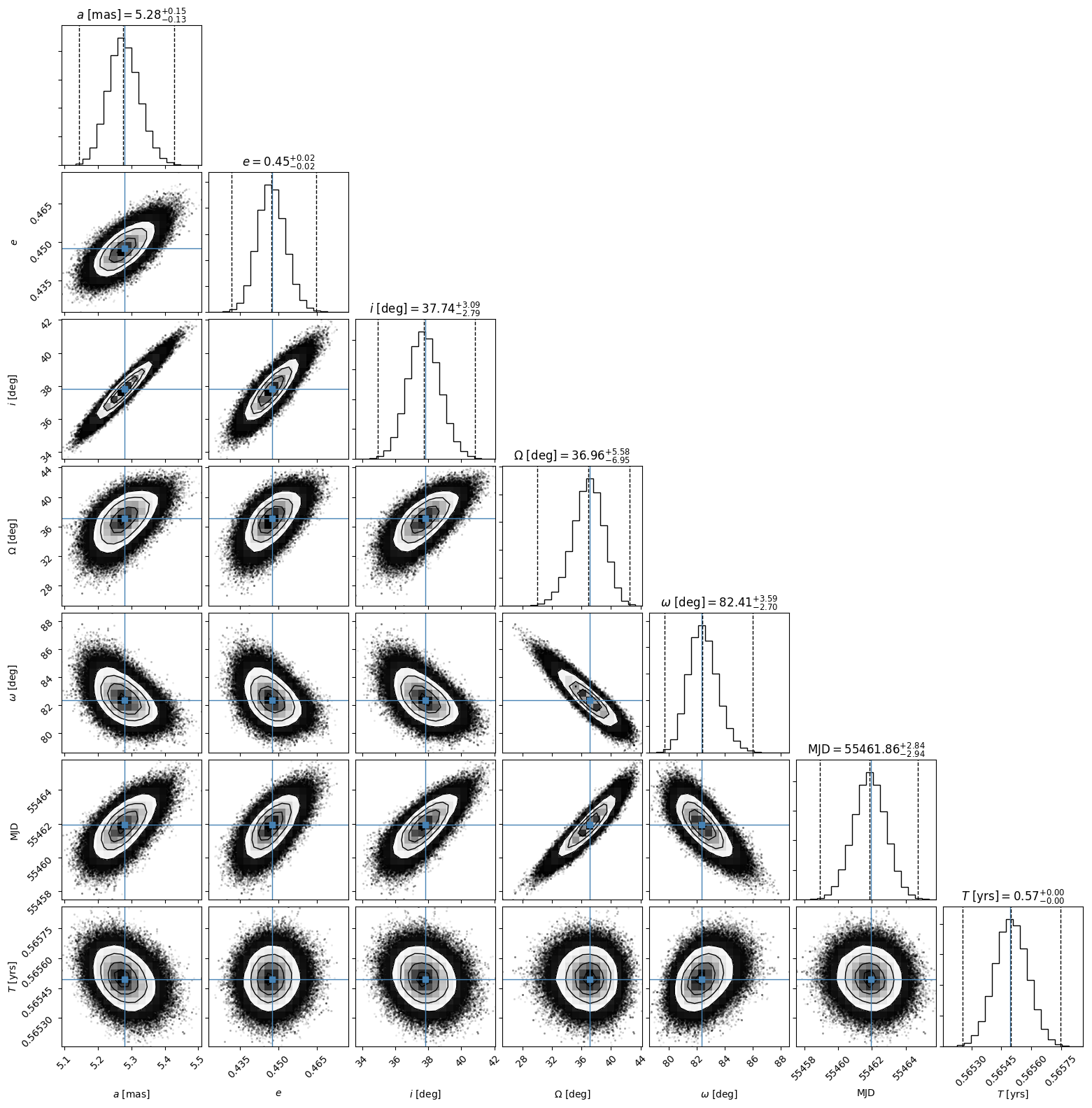}
    \caption{MCMC marginal posterior distributions of the fitted orbital parameters for all nine epochs. The errors are the 3$\sigma$ confidence intervals. The blue line marks the values of the best fit; the inner dashed line is the median of the distribution, and the outer dashed lines contain the 3$\sigma$ confidence intervals.}
    \label{fig:corner_a2b1cd1d05fgh28h30_truths_e02}
\end{figure*}

Finally, we refined the orbital solution by fitting the secondary position for all nine epochs together. The final orbital parameters and the $3\sigma$ confidence interval were determined by an MCMC estimation\footnote{The difference between this and the first MCMC fit is that, for the first, a large number of walkers (500) are randomly distributed over a large range of the possible parameter space, while for the second, the initial distribution of the walkers is drawn very closely (i.e., in an interval smaller than the resulting confidence interval) around the previously found best fit.}; the result is shown in Table~\ref{tab:orbit_result}, the derived orbit in Fig.~\ref{fig:orbit_all9_e02}, and the MCMC marginal posterior distributions of the parameters in Fig.~\ref{fig:corner_a2b1cd1d05fgh28h30_truths_e02}. Interestingly, the astrometric residuals of the best-fit orbit at the three GRAVITY epochs are found to be less than 40\,$\mu$as, in agreement with our estimate of the systematic errors in Sect.~\ref{sec:positions}.
\begin{table}
        \caption{Results of the orbit fit.}
        \centering
        \begin{tabular}{lrl}
                \hline \hline
                Parameter                      & Value   & $3\sigma$ uncertainty \\ \hline 
                $a$ [mas]                      & 5.28    & $^{+0.15}_{-0.13}$\vspace{0.1cm} \\
                $a$ [au]                       & 1.01    & $^{+0.03}_{-0.02}$\vspace{0.15cm} \\
                $T$ [days]                     & 206.55  & $^{+0.09}_{-0.09}$\vspace{0.15cm} \\
                $e$                            & 0.45    & $^{+0.02}_{-0.02}$\vspace{0.15cm} \\
                $i\,[^{\circ}]$                & 37.7    & $^{+3.1}_{-2.8}$\vspace{0.15cm} \\
                $\Omega\, [^{\circ}]$          & 37.0    & $^{+5.6}_{-6.9}$\vspace{0.15cm} \\
                $\omega\,[^{\circ}]$           & 82.4    & $^{+3.6}_{-2.7}$\vspace{0.15cm} \\
                $t_{\mathrm{p}}$ [MJD]         & 55461.9 & $^{+2.8}_{-2.9}$\vspace{0.15cm} \\
                $M_{\mathrm{tot}}$ [$M_\odot$] & 3.20    & $^{+0.30}_{-0.25}$\vspace{0.15cm} \\
                $\chi^2_{red}$                 & 2.03\\
                \hline
        \end{tabular}
        \tablefoot{The parameters are semimajor axis $a$, eccentricity $e$, inclination $i$, longitude of the node\tablefootmark{*} $\Omega$, argument of periastron $\omega$, time of periastron $t_{\mathrm{p}}$, and period $T$. The combined mass of the stars, $M_{\rm{tot}}$, is calculated assuming a distance of $d=191.0$~pc. ($^{\ast}$) The position angles of the ascending node and descending node are not distinguishable; this is the one that is $<$180$^{\circ}$.
        }
        \label{tab:orbit_result}
\end{table}
\subsection{Image reconstruction}\label{sect:imagerecon}
For two of the GRAVITY epochs, namely 2019-Jun-20 and 2020-Jan-30, the uv-plane is sampled well enough (see Fig.~\ref{fig:data2}) to attempt a first-order model-independent image reconstruction for a simple binary object. The primary goal is to verify the consistency of the retrieved positions and flux ratios with the model fitting approach.

\begin{figure*}
    \begin{subfigure}{0.44\textwidth}
        \includegraphics[width=\linewidth]{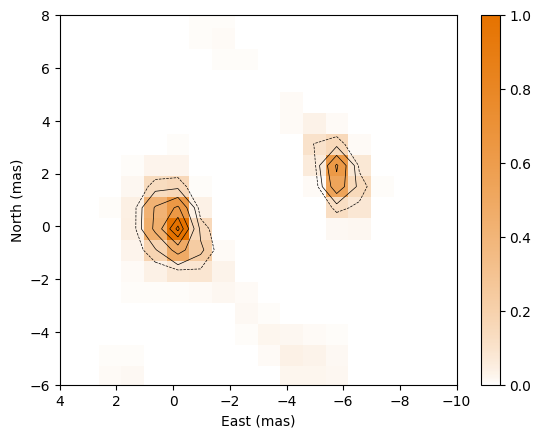}
        \caption{2019-Jun-20}
    \end{subfigure}
    \hspace*{\fill}
    \begin{subfigure}{0.44\textwidth}
        \includegraphics[width=\linewidth]{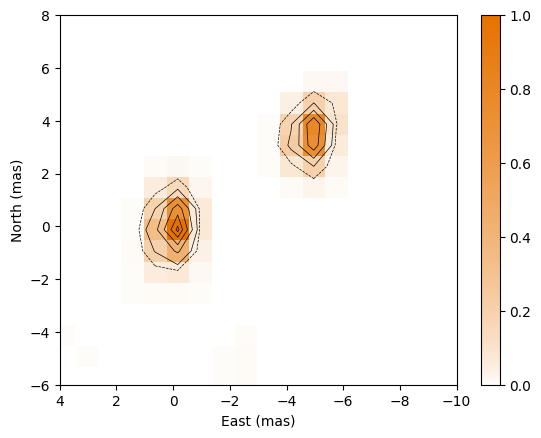}
        \caption{2020-Jan-30}
    \end{subfigure}
    \caption{SQUEEZE image reconstruction with a pixel scale of 0.8\,mas/px for runs G (2019-Jun-20) and I (2020-Jan-30). The solid contours represent the $95\%$, $80\%$, $60\%$, and $20\%$ flux levels. The dashed contour (10\% of the maximum flux) represents the 3$\sigma$ detection limit from the background artifacts. 
    Between both epochs lie 1.08 orbit periods. 
    }
    \label{fig:squeeze2019}
\end{figure*}

For this purpose, we applied the SQUEEZE algorithm \citep{2010SPIE.7734E..2IB}. For the results shown in Fig.~\ref{fig:squeeze2019} and reported in Table~\ref{tab:position_fits}, we used no initial image, which, in the case of SQUEEZE, implies starting with all of the flux concentrated in the central pixel of the grid. The pixel scale is usually defined by the resolution of the interferometer and is thus dependent on the longest projected baseline. Using oversampling by a factor of two  ($\Delta\theta\lesssim\lambda$/($4B_{max}$), according to \cite{2017JOSAA..34..904T}) results in pixel scales of 0.9\,mas/px and 1.1\,mas/px for the 2019 and 2020 data, respectively. We thus chose a pixel scale of 0.8\,mas/px for both epochs. The edge length of the squared field of view is 30.4\,mas and was chosen to be approximately three times larger than the main extent of the reconstructed object. Except for the field of view regularization, which keeps the flux centroid in the middle of the image, no other regularization was used.

Other combinations in addition to the above settings were tested. These include starting from a random image, pixel scales of 0.5\,mas/px and 0.2\,mas/px, a field of view of $60 \times 60$\,mas$^2$, and using additional regularization functions, such as the total variation and the L0 sparsity norm. None of the alternative reconstruction results deviate strongly from the one described above when the following criteria are fulfilled: They have the same relative positions (within the errors), they have similar flux ratios, there are no additional substructures, and the two main features have a similar shape. The results of the tests for three different pixel scales are illustrated in Fig.~\ref{fig:img-rec-app} and clearly show a similar trend in terms of astrometry and relative flux.

The result of the image reconstruction shows two elongated Gaussian shapes with changing position angles, as seen in Fig.~\ref{fig:squeeze2019}. The orbital motion is clearly detected. The flux ratio between the two components appears to change noticeably between the two epochs. By fitting an elongated Gaussian to both bright sources, we derived the relative position and flux ratio (as the ratio of the height of the Gaussians) as well as the corresponding uncertainties, which are reported in Table~\ref{tab:position_fits}. The uncertainties are the uncertainties of the Gauss fits, and further uncertainties coming from the image reconstruction process are not taken into account. We find that the position of the secondary measured in the reconstructed image is consistent with the results of the parametric fit within 30\,$\mu$as, which indicates the robustness of the result regarding this quantity. Similarly, the derived flux ratios are also in very good agreement.

We note that the elongation of the Gaussian shapes and their changing position angles, as well as the additional fainter structures in the reconstructed image, most likely result from the incomplete and irregular uv-coverage, which may produce image artifacts. Therefore, considering the limited spatial resolution in comparison to the size of any extended circumstellar emission, further interpretation of the relative elongation of the reconstructed Gaussian blobs in terms of the spatial properties of the circumstellar disks has not been the aim of this section.
%
\begin{figure}
    \centering
        \includegraphics[width=0.92\columnwidth]{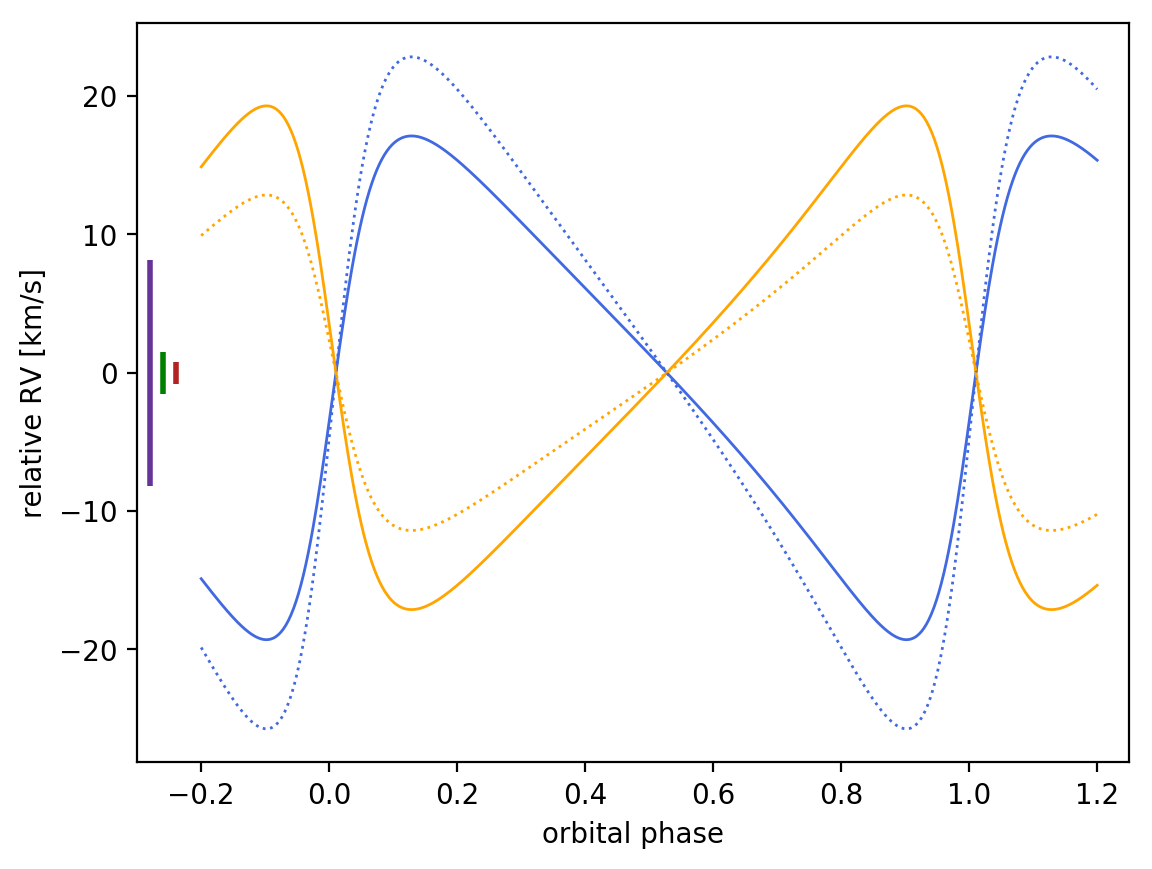}
    \caption{Relative (with respect to barycenter) RVs as derived from our orbital solution for the cases of $q=1$ (solid) and $q=0.5$ (dotted). The RV of the primary is in orange, and the companion is in blue. Depicted on the left are the spectral resolutions of X-SHOOTER, CRIRES, and ESPRESSO in purple, green, and red, respectively. We note that the RV precision is much more accurate.}
    \label{fig:RV}
\end{figure}
\section{Discussion}
\label{sec:results}
\subsection{Observed properties of WW\,Cha}
The squared visibilities and closure phases measured with the three VLTI instruments clearly show a well-resolved source with clear evidence of asymmetry in the continuum emission. As seen in Fig~\ref{fig:data2}, the sinusoidal shape of the squared visibilities as a function of the spatial frequencies points to a general trend expected for a binary star. The closure phase signal is strong, exceeding 100$^{\circ}$ on many occasions. The resulting astrometry evidences a separation varying from 4.6\,mas to 6.3\,mas. With the adopted model, both the H and K band photometry is consistent with a flux ratio between 0.5 and 0.9. The contribution of the extended emission is significant, with a relative contribution to the total flux of 13\% to 24\% (cf. Table~\ref{tab:position_fits}).
Complementing the results of \citet{2015A&A...574A..41A} with new observations, we propose, for the first time, an orbital solution for this young T Tauri system. 
A period of $T=206.55^{+0.09}_{-0.09}$~days and a semimajor axis of $a = 1.01^{+0.02}_{-0.02}$~au place WW\,Cha in a range of separation that has been poorly covered by previous large imaging surveys in Chamaeleon \citep{Lafreniere2008,Daemgen2016}.

 The relatively short period paired with the available knowledge of the full orbital parameters make WW\,Cha easy to survey.\ It is thus a prime target for testing theories of dynamical interactions between the binary star and its environment.\vspace{0.2cm}\\
%
\subsection{Dynamical mass}
The dynamical mass of the system can be self-consistently derived from the knowledge of the orbital solution and is found to be $M_{\rm tot}=3.20^{+0.30}_{-0.25}\,M_{\odot}$ (3$\sigma$ confidence level). 
This classifies the object as an intermediate-mass T\,Tauri binary star, assuming a mass ratio of unity and following \cite{Calvet2004}, who define intermediate-mass T\,Tauri stars with 1\,$M_\odot$\,$\leq$\,$M$\,$\leq$\,5\,$M_\odot$.
The mass ratio cannot be measured directly from these data and will become accessible from spectroscopic RV measurements combined with the knowledge of the orbit inclination reported here.

The RV of WW\,Cha as predicted from our orbital solution ranges from -19.3\,km\,s$^{-1}$ to 17.1\,km\,s$^{-1}$ in the case of an equal mass binary ($q$=1), and from -25.8\,km\,s$^{-1}$ to 22.9\,km\,s$^{-1}$ for the lower-mass companion for $q$=0.5. The maximum RV separation of WW\,Cha is 38.5\,km\,s$^{-1}$ (independently of the mass ratio; see Fig.~\ref{fig:RV}).

\citet{2012ApJ...745..119N} monitored the RV of WW\,Cha (among 211 other T\,Tauri stars) for four epochs between February and December 2006.\ They measured a relatively high weighted RV standard deviation of 6\,km\,s$^{-1}$ but did not consider WW\,Cha as a candidate for binarity due to the relatively high systematic noise of 4\,km\,s$^{-1}$. WW\,Cha was also not detected as a double-peaked spectroscopic binary (SB2) by the same authors, although they report $\sim$10\,km\,s$^{-1}$ as a lower limit for a detectable velocity separation, which is about twice the sampling interval. One reason could be that WW\,Cha was observed within epochs with lower RV separation\footnote{This is not tested here since the observation epochs are not accessible.}.

Based on Fig.~\ref{fig:RV}, the required spectral resolution to resolve WW\,Cha as an SB2 binary would be $\sim$\,10\,km\,s$^{-1}$ (i.e., $R$\,$\sim$\,30,000). Even twice this resolution is accessible with instruments 
such as CRIRES and ESPRESSO, with $R$\,=\,60,000 and $R$\,=\,140,000, respectively. Using our orbital solution and derived expected RVs will help in finding the binary signal in spectroscopic data as well as in planning spectroscopic observations at the best suitable epochs.
\vspace{0.2cm}\\
\subsection{Stellar parameters}
In order to further constrain the stellar properties of the individual components, we adopted the classical approach that consists in comparing the mass and temperature to PMS evolutionary models. We used and compared the Siess tracks \citep{2000A&A...358..593S} and the Pisa tracks \citep{Tognelli2011}, accounting for a marginal subsolar metallicity of Cha\,I \citep{James2006} in order to explore the range of possible mass ratios $q$ for the WW\,Cha system. 
An alternative way would be to analyze each component separately, making use of the individual H and K band photometry that was derived by comparing the 2MASS \citep{2006AJ....131.1163S} total magnitudes with the flux ratio $f_{comp}/f_{prim}$ (Table \ref{tab:position_fits}). However, the flux ratio as derived from our interferometric fits is not directly comparable to the intrinsic stellar flux ratio.
This is because it is likely that part of the H and K band flux originates from circumstellar warm and hot material.

We consider two cases for the age of WW\,Cha, namely the median age of Chamaeleon\,I of 2\,Myr, as previously estimated at 160\,pc by \citet{2007ApJS..173..104L}, and the age of 1\,Myr that was estimated when revising the distance to 190\,pc. 
We also considered a range of effective temperatures of the primary -- which dominates the system photometry -- from 4350\,K (K5) to 5110\,K (K0), as reported by \citet{2007ApJS..173..104L} and \citet{2016A&A...585A.136M} through spectral fitting in the red (i.e., 640-880\,nm) and the blue (i.e., 340-460\,nm) optical spectrum. Regarding the large difference in spectral type between these two sources, we note that \citet{Pascucci2016} favored the earliest spectral type, citing the lack of good temperature diagnostics in the low-resolution spectra from \citet{2007ApJS..173..104L}.

\begin{figure}
    \centering
        \includegraphics[width=\columnwidth]{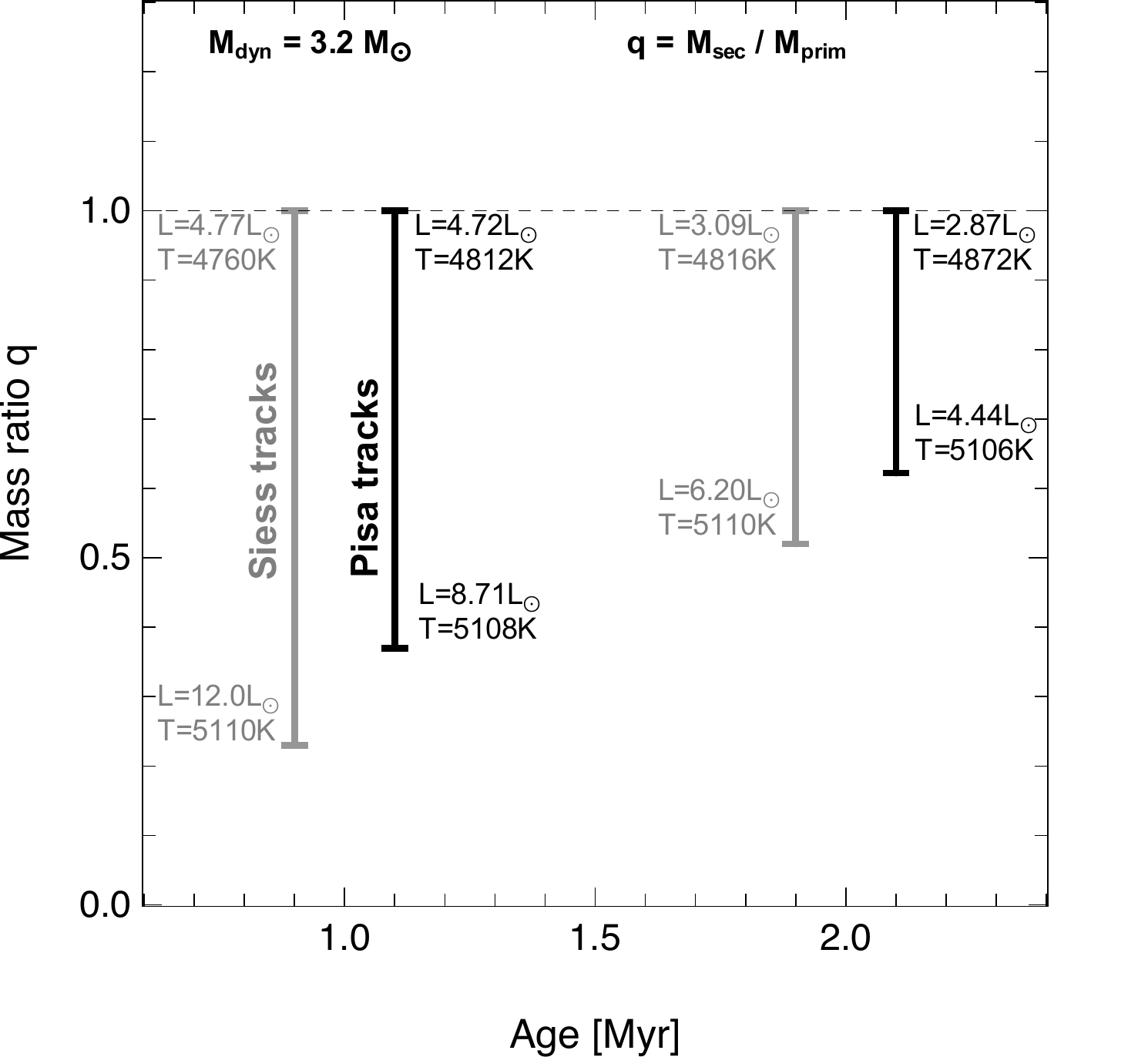}
    \caption{Derivation of the stellar mass ratio in WW\,Cha via comparison with evolutionary models. The gray curve uses the Siess tracks, and the black curve considers the Pisa tracks. The two median ages considered are 1 and 2\,Myr. For each age, two vertical bars are plotted apart for better readability. The derived temperature and luminosity are only reported for the primary component for the two extreme values of the mass ratio. 
    }
    \label{fig:HRD-tracks}
\end{figure}

With this information and the knowledge of the dynamical mass, we explored the range of the $q$ parameter, accounting for a minimum mass of the primary of 1.6\,$M_{\odot}$. Figure~\ref{fig:HRD-tracks} shows the expected differences in the derived mass ratio with the considered evolutionary model, although the resulting range of possible mass ratios for the two estimated ages remains qualitatively comparable. At the younger age of $\sim$1\,Myr, our derived dynamical mass is compatible with a mass ratio $q$ that is between $\sim$0.3 and 1; it is limited to $\sim$0.5\,--\,1 at the more evolved age of $\sim$2\,Myr.

\
This result shows that inferring stellar parameters in a binary system with an independently known total mass is possible. 
However, this approach is well known to be critically dependent on reliable estimates of the effective temperature and age, which suggests that a more accurate estimate of the mass ratio needs, in the end, to be robustly complemented by spectroscopic RV measurements. 
\vspace{0.2cm}\\
\subsection{Interaction with disk(s)}
The spectral energy distribution of WW\,Cha exhibits a large infrared excess that is typically associated with disk-like emission in the system. Since WW\,Cha is found to be a close ($\sim$1\,au) binary, we expect strong dynamical effects on the disk environment due to the tidal forces.

%
%
The observations that can best be compared to the orbit are observations of the circumbinary disk around WW\,Cha obtained with VLT/SPHERE \citep{2020A&A...633A..82G} and ALMA \citep{Pascucci2016}. 
While in the SPHERE image the innermost 35~au are blocked by the coronagraph, the image shows a large outer disk with spiral structures that extend to at least 100~au as well as filaments that reach out as far as a few hundred astronomical units. The disk inclination is not clearly determined from SPHERE due to the presence of spiral structures; however, it is considered to be moderate by the authors and seems to be in coarse agreement with our measured value, $i=37.7^{+3.1}_{-2.8}$ degrees of the orbit. A much better estimate of the disk inclination comes from the ALMA observations \citep{Pascucci2016}, which fit the submillimeter emission with an elliptical Gaussian with a full width at half maximum  of $0.56 \times 0.44$\,arcsec. This translates into an inclination of $i=38.2$\,degrees, which is remarkably close to the orbit inclination, although no error bars are reported. 
\citet{2020A&A...633A..82G} estimated the position angle of the disk to be $\sim$\,50$^{\circ}$, which is comparable to the longitude of the ascending node of the orbit, $\Omega$=37.0$^{+5.6}_{-6.9}$ degrees. Combined with the ALMA inclination, this would result in a mutual disk-binary inclination of $\theta$\,$\sim$\,8$^{\circ}$ (cf. \citet{Czekala2019}, Eq.\,(1)), and thus hints at a coplanarity between the circumbinary disk and binary orbit.

The disk's spiral features may be interpreted as tidal disruption that is induced by the binary in the inner disk and is propagating outward. This assumption is supported by the clockwise wrapping of the spiral structures, which is expected for a binary in counterclockwise rotation and has been predicted in this form by simulations (e.g., by \citet{2002A&A...387..550G}, \citet{2015MNRAS.448.3545D}, \citet{2016ApJ...827...43M}, \citet{2020ApJ...889..114M}). 
With our new description of WW\,Cha's orbit, future hydrodynamic simulations, tailored to these findings, will explore the finer dynamical interactions between the disk and the close binary, while new ALMA observations will explore the inner regions of the circumbinary disk (e.g., \citet{Kurtovic2018}).

As opposed to GRAVITY, SPHERE does not deliver sufficient spatial resolution to explore the immediate environment of the binary components themselves. Although a detailed morphological study of potential circumstellar disk components is beyond the scope of this work, a simple photometric analysis of our results in this context is still meaningful. 
Assuming a mass ratio $q$=1, the theoretical H and K band stellar magnitudes of the individual components can be derived from the stellar parameters and extinction. At a distance of 191\,pc, the apparent magnitudes of both equal-mass components are $m_H$=8.78 and $m_K$=8.26 for $A_V$\,$\sim$\,4.8 \citep{2007ApJS..173..104L} at 1\,Myr, with an uncertainty of 0.02 magnitude due to the error on the distance. The comparison with the estimated magnitudes in Table~\ref{tab:position_fits} shows that, particularly in the K band, the measured flux is higher than what is expected from a pure photospheric emission. Since the near-infrared excess is unlikely to originate in the inward truncated circumbinary disk, it is plausible to interpret the detected excess at the position of each component as surviving circumstellar disk(s). It should be noted that the same conclusion can be drawn when considering a smaller mass ratio. For $q$=0.4, the primary and secondary components would have, respectively, $m_H$=\{8.21, 9.32\} and $m_K$=\{7.73, 8.78\}.
It is then interesting to notice that, in the context of our understanding of disk dispersal in a dynamically constrained environment, the small semimajor axis and relatively high eccentricity of the orbit would not yet have resulted, at this young age, in an efficient dynamical dispersal of the existing circumstellar disks. The picture of WW\,Cha as a single-star accretor by \cite{Daemgen2016} needs to be revised.
\vspace{0.2cm}\\
%
\subsection{Variability}
As a young PMS binary star, WW\,Cha shows emission lines both in H$\alpha$ \citep{Robberto2012} and in Br$\gamma$ \citep{Daemgen2016}, which is interpreted as a signpost of gas accretion \citep{McCabe2006}. In this context, the close binary nature of WW\,Cha leads to interesting questions regarding the connection between source variability and orbital events. For instance, in the case of the close spectroscopic T\,Tauri binary DQ\,Tau ($P$=15.8\,d, $a$=0.13\,au, $e$=0.57, \citet{2016ApJ...818..156C}), which is surrounded by a circumbinary disk, \citet{2019ApJ...877...29M} have shown that a periodic, sudden, and very short-term increase in flux around the time of closest approach is explained by the disruption of the circumstellar gas and the subsequent accretion burst due to the rapid fall of matter onto the central stars. Although at a different separation scale, the potential circumstellar disks around each component may dynamically interact and trigger variability effects that could be monitored against the well-constrained time of peri-passage. 
With a derived eccentricity of $e$=0.45, the separation between the two components will vary by a factor of $\sim$2.5 on a $\sim$100\,d timescale, which offers an excellent opportunity to test such effects. Interestingly, with the current data, we observe for the K band a significant change in the flux ratio between the two components, from $\sim$0.5 (Run G) to $\sim$0.8 (Run H and I), over a $\sim$1 orbit period, which occurs close to two successive passages to apoastron. Finally, considering the expected small sizes of possibly surviving circumstellar disks, witnessing the accretion of matter directly from the circumbinary disk, similarly to what was reported at a different spatial scale by \citet{Alves2019} in [BHB2007]\,11, cannot be ruled out. Further follow-up observations will help us to clarify the origin of this variation in the flux contrast.
%
%
\section{Conclusion}
\label{sec:conclusion}
In this work, we report a detailed astrometric and photometric study of the PMS binary star WW\,Cha (located at a DR2 distance of 191\,pc) using new near-infrared interferometric data obtained with the VLTI. We obtain the following results:
\begin{itemize}
\item We robustly constrain the orbital motion of the close pair and derive, with high accuracy, the set of orbital parameters. WW\,Cha, with $\sim$1\,au separation,  presents a relatively eccentric orbit ($e$=0.45). The mutual inclination between the binary orbit and the circumbinary disk is low ($\theta$\,$\sim$8$^{\circ}$). 
WW\,Cha is an illustration of the fact that, although disks in binary systems with periods $P$\,$>$\,30\,d or $e$\,$>$0.2 do not necessarily favor coplanarity \citep{Czekala2019}, close alignment with the system's orbit cannot be excluded.\vspace{0.2cm}
\item The total stellar mass is estimated to be 3.2\,$M_\odot$, a factor of $\sim$2 more massive than what had been reported from spatially unresolved photometric and spectroscopic analysis that assumed the case of a single star \citep{Manara2016}. Accounting for the range of the spectral types that exist in the literature, we estimate a plausible mass ratio between $\sim$0.5 and 1, depending on the age of the system.  
\item Assuming a geometrical model of the binary composed of two point sources and a resolved extended emission, we measure the flux ratio of the two components and derive the individual H and K band magnitudes. We compare them to the theoretical pure photospheric magnitudes and find a magnitude excess of $\Delta K$\,$\sim$1-1.5, which could be explained by the presence of dynamically truncated circumstellar disks.
\end{itemize}
%
With the knowledge of the orbit inclination, follow-up high-resolution spectroscopic observations will help to precisely determine the individual stellar masses using RV measurements. 
With the well-constrained orbital and physical parameters, and thanks to the extended wavelength coverage of high-angular-resolution datasets (SPHERE, VLTI, ALMA), WW\,Cha has become an ideal benchmark for the study of disk evolution in close $\sim$1\,yr period binary stars, from both an observational and a theoretical standpoint.

\begin{acknowledgements}
We thank the anonymous referee for assessing the quality of this work. 
This work is based on observations made with ESO Telescopes at the La Silla Paranal Observatory under programme IDs 094.C-0884, 94.C-0884, 082.C-0920, 098.C-0334, 098.C-0334, 0103.C-0347, 0104.C-0567, 0104.C-0567.
F.E. is supported by the University of Cologne and the Bonn-Cologne Graduate School of Physics and Astronomy (BCGS). F.E. kindly acknowledges the financial support of this work by the EII - Fizeau Program funded by WP11 of OPTICON/H2020 (2017-2020, grant agreement 730890).
J.S.B. acknowledges the full support from the UNAM PAPIIT project IA 101220. 
R.G.L. acknowledges support from Science Foundation Ireland under Grant No. 18/SIRG/5597.
A.C.G. has received funding from the European Research Council (ERC) under the European Union’s Horizon 2020 research and innovation programme (grant agreement No. 743029).
A.A. and P.G. were supported by Funda\c{c}\~{a}o para a Ci\^{e}ncia e a Tecnologia, with grants reference UIDB/00099/2020 and SFRH/BSAB/142940/2018.
This research has made use of the  \texttt{AMBER data reduction package} and \texttt{PIONIER data reduction package} of the Jean-Marie Mariotti Center\footnote{Available at http://www.jmmc.fr/amberdrs and at http://www.jmmc.fr/pionier}. This research has made use of the Jean-Marie Mariotti Center OiDB service available at http://oidb.jmmc.fr. 
This research has made use of the Jean-Marie Mariotti Center \texttt{LITpro} service co-developed by CRAL, IPAG and LAGRANGE \footnote{LITpro software available at http://www.jmmc.fr/litpro}.
\end{acknowledgements}

\bibliographystyle{aa} 
\bibliography{WW_Cha_Orbit} 

\appendix
\section{VLTI data}
\begin{table*}[h!]
        \caption{Observation log.}
        \centering
        \begin{tabular}{lcllcccccc}
                \hline \hline
                Date & MJD       & Run &Instrument   & Configuration  & Calibrator (diam. [mas]) & Seeing ($^{\prime\prime}$)& Airmass & $\tau_0$ [ms] &  
                \\ \hline
                10/2/11 & 55602.20 & A & PIONIER     & A0-G1-I1-K0        & HIP 55237 ($0.207$) & 0.60 -- 0.65 & 1.68 & 7 & 
                \\
                6/3/12 & 55992.14 & B & PIONIER     & A1-G1-I1-K0        & HIP 56876 ($0.253$) & 1.27 & 1.67 & 4 &
                \\
                   &        &    &           &                     & HD 54452 ($0.351$) & & & &  \\ 
                2/7/12 & 56110.02 & C & PIONIER     & A1-G1-I1-K0        & HD 99015 ($0.239$) & 0.81 -- 0.88 & 1.86 -- 1.87 & 4 &
                \\
                5/2/15 & 57058.13 & D & PIONIER   &   D0-G1-H0-I1       & HD 94246 ($0.140$) & 0.67 -- 0.92 & 1.81 -- 2.06 & 9 -- 14 & \\
                &           &    &           &                     & HD 96494 ($0.179$) & & & &  \\ 
                &          &    &           &                     & HD 94189 ($ 0.124 $) & & & &  \\
                &          &    &           &                     & HD 89591 ($ 0.160  $) & & & &  \\
        13/2/15 & 57066.23 & E & PIONIER     & A1-G1-K0-J3    & HD 89591 ($ 0.160 $) & 1.04 -- 1.08 & 1.64 & 7 -- 8 &\\
                9/3/17 & 57821.35 & F  & AMBER        & U2-U3-U4        & HD 99015 ($0.239$) & 0.35 -- 0.37 & 1.90 & 8 -- 9 &  \\
                20/6/19 & 58654.01 & G  & GRAVITY     & U1-U2-U3-U4       & HD 99556 ($0.221$) & 0.39 -- 0.59 & 1.72 -- 1.82 & 6 -- 8 &  \\
                28/1/20 & 58876.39 & H  & GRAVITY       & D0-G2-J3-K0       & HD 99556 ($0.221$) & 0.80 & 1.69 -- 1.70 & 4 & \\
                30/1/20 & 58878.34 & I  & GRAVITY     & D0-G2-J3-K0       &  HD99556 ($0.221$) & 0.55 -- 1.20 & 1.62 -- 1.73 & 3 -- 6 & \\
        &                  &    &           &                     & HD 98142 ($0.038$)& &   \\ 
                \hline
        \end{tabular}
        \tablefoot{The date format is day-month-year. Runs A, B, and C are not present in the ESO archive. For the remaining runs, the corresponding IDs are: D=094.C-0884, E=094.C-0884, F=082.C-0920, G=098.C-0334, H=098.C-0334, I=0103.C-0347, J=0104.C-0567, and K=0104.C-0567. Runs A, B, and C are from \citet{2015A&A...574A..41A}. The error bars on the diameter of the calibrators span from $\pm$0.001 to $\pm$0.009. }
        \label{tab:obs_log}
\end{table*}
\begin{figure*}[h]
   \centering
   \includegraphics[width=0.96\textwidth]{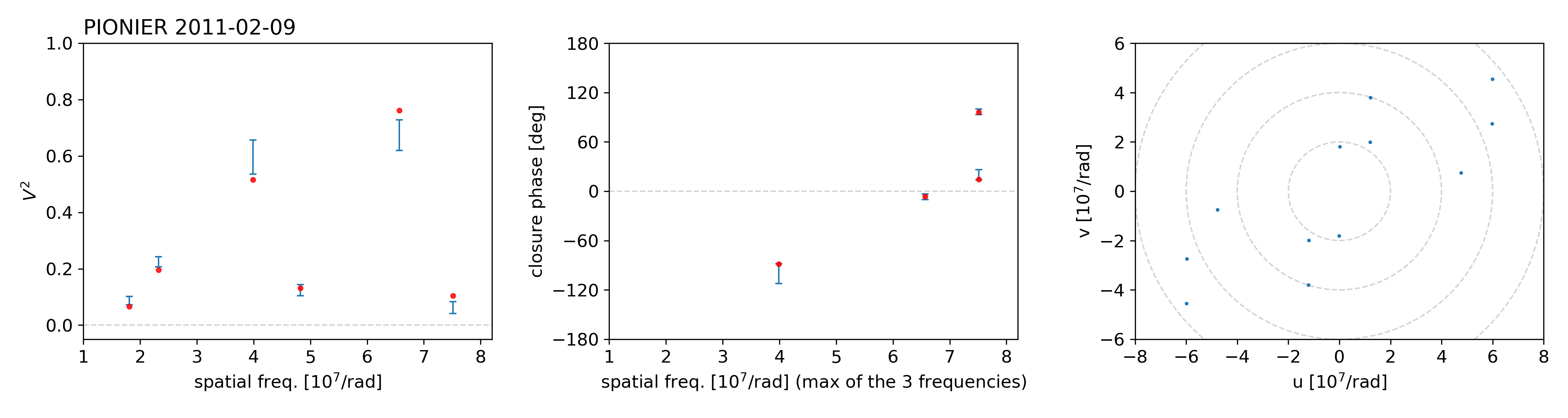}\\
   \includegraphics[width=0.96\textwidth]{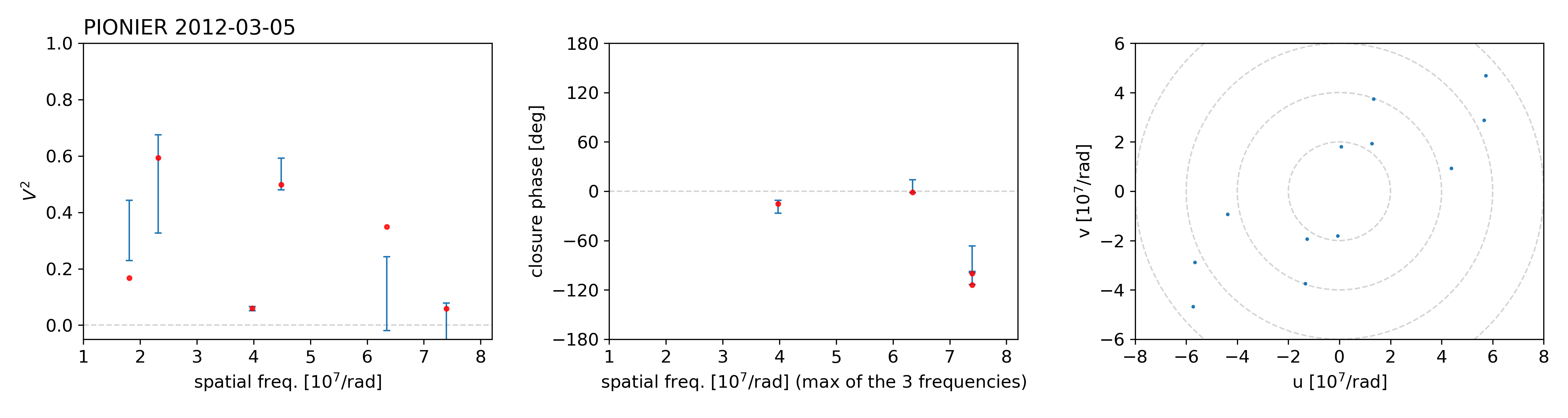}\\
   \includegraphics[width=0.96\textwidth]{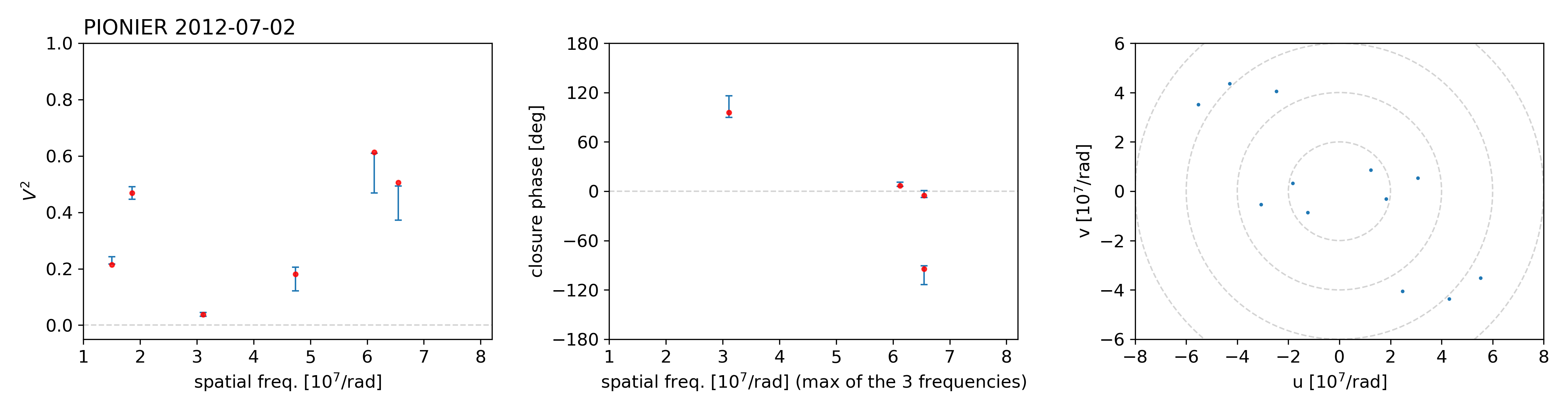}\\
   \includegraphics[width=0.96\textwidth]{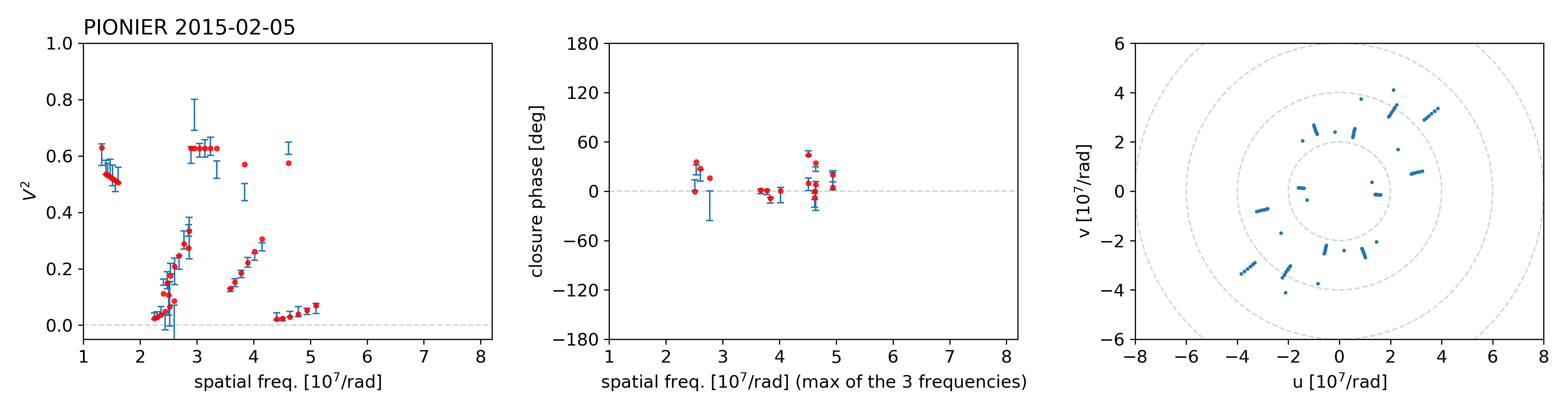}\\
   \includegraphics[width=0.96\textwidth]{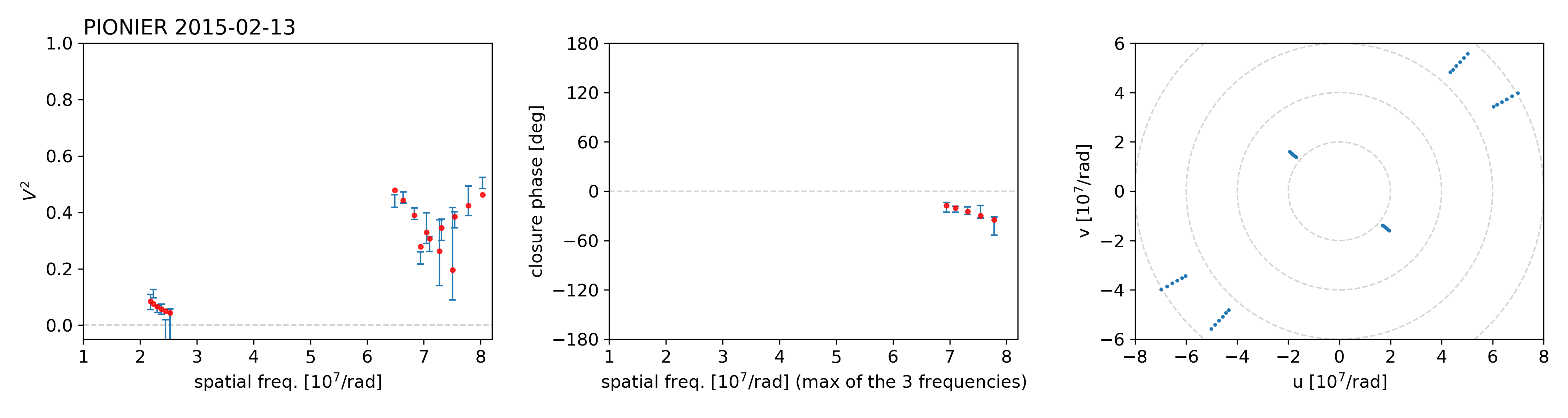}
   \caption{PIONIER interferometric data (blue points with error bars) and the best-fit result (red filled circles) for each epoch. From left to right are shown the squared visibilities, the closure phases, and the uv-plane coverage.}
   \label{fig:data1}%
\end{figure*}
\begin{figure*}
   \centering
   \includegraphics[width=0.96\textwidth]{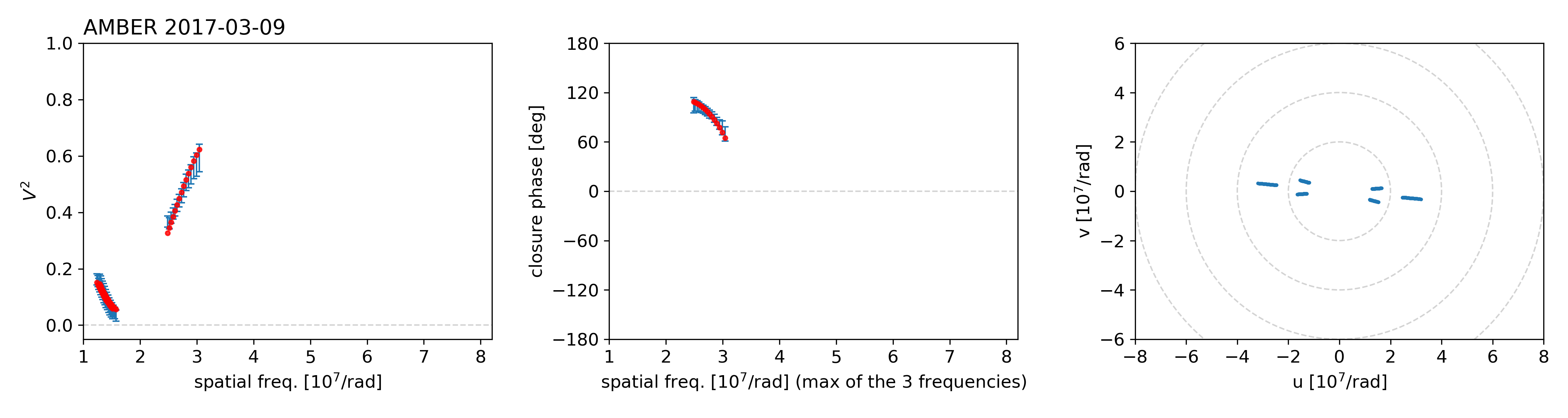}\\
   \includegraphics[width=0.96\textwidth]{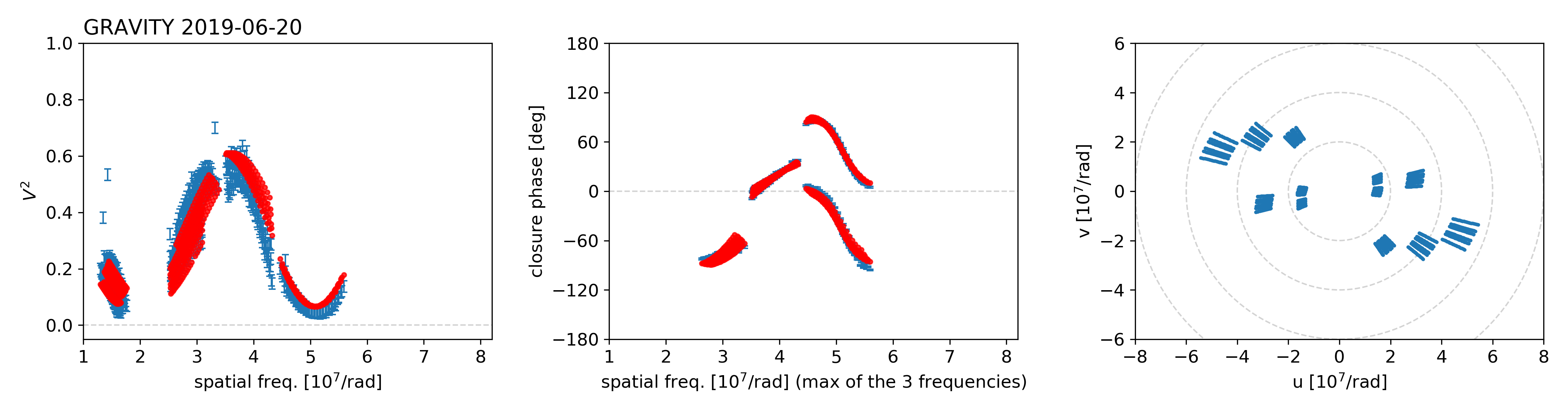}\\
   \includegraphics[width=0.96\textwidth]{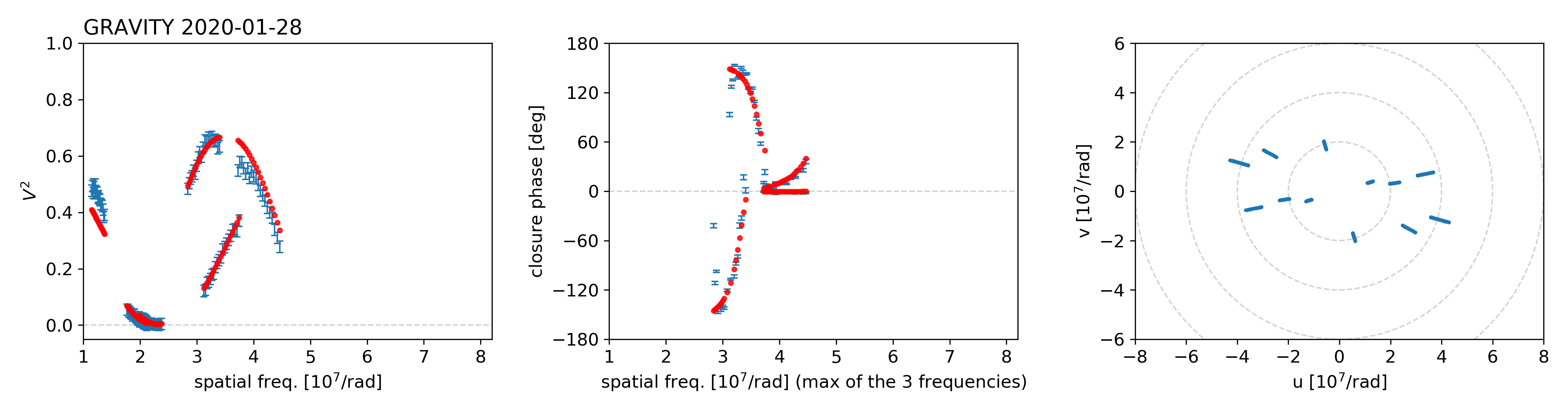}\\
   \includegraphics[width=0.96\textwidth]{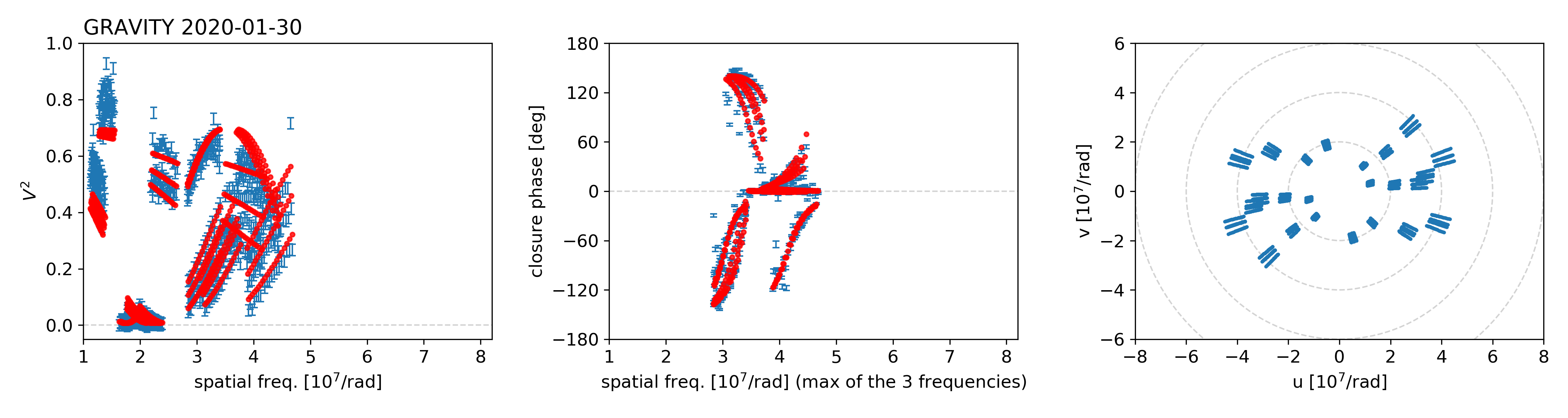}
   \caption{Same as Fig. \ref{fig:data1} but for AMBER and GRAVITY.}
   \label{fig:data2}%
\end{figure*}
\newpage
\clearpage
\section{Minimization maps}
\begin{figure*}[b]
\begin{subfigure}{0.32\textwidth}
        \includegraphics[width=\linewidth]{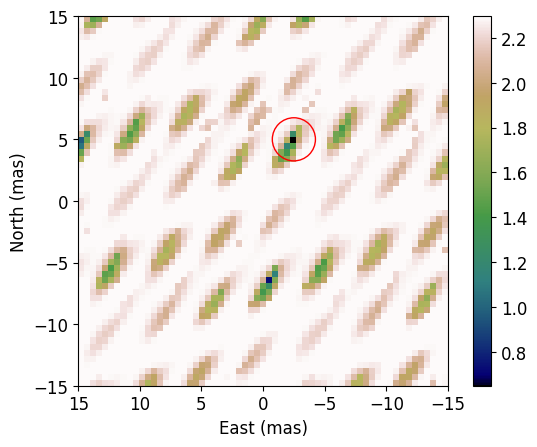}
    \caption{2011-Feb-10 (A)}
    \label{fig:chi2map_A}
\end{subfigure}
\hspace*{\fill}
\begin{subfigure}{0.32\textwidth}
        \includegraphics[width=\linewidth]{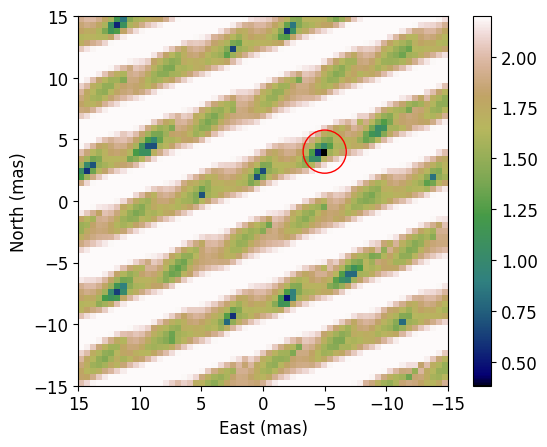}
    \caption{2012-Mar-06 (B)}
    \label{fig:chi2map_B}
\end{subfigure}
\hspace*{\fill}
\begin{subfigure}{0.32\textwidth}
        \includegraphics[width=\linewidth]{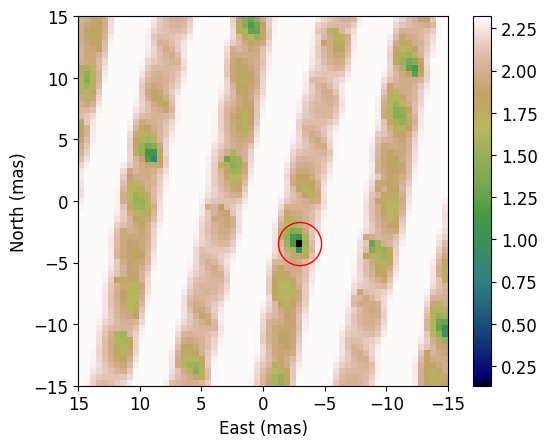}
    \caption{2012-Jul-02 (C)}
    \label{fig:chi2map_C}
\end{subfigure}
\\
\begin{subfigure}{0.32\textwidth}
        \includegraphics[width=\linewidth]{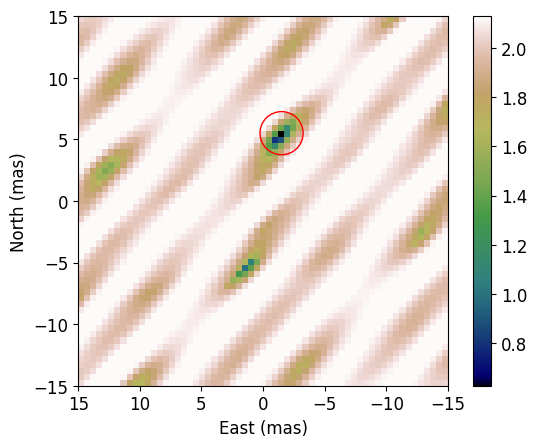}
    \caption{2015-Feb-05 (D)}
    \label{fig:chi2map_D}
\end{subfigure}
\hspace*{\fill}
\begin{subfigure}{0.32\textwidth}
        \includegraphics[width=\linewidth]{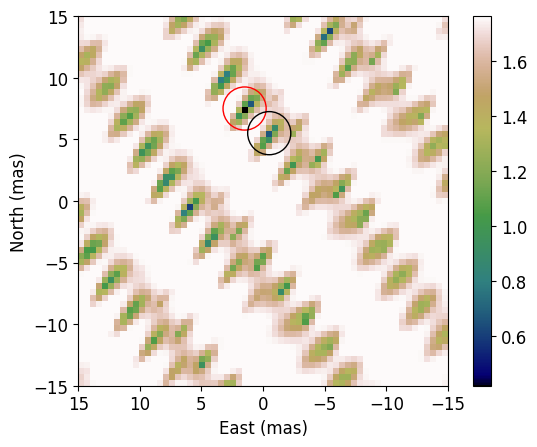}
    \caption{2015-Feb-13 (E)}
    \label{fig:chi2map_E}
\end{subfigure}
\hspace*{\fill}
\begin{subfigure}{0.32\textwidth}
        \includegraphics[width=\linewidth]{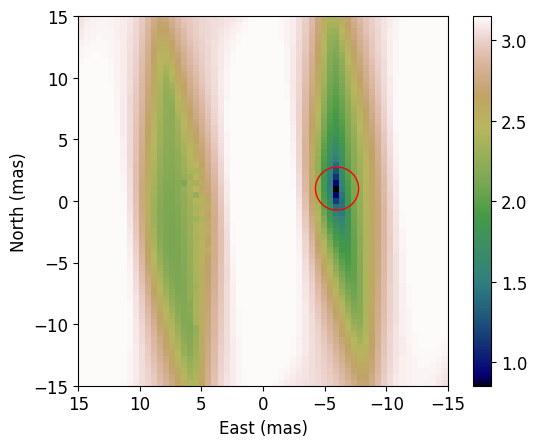}
    \caption{2017-Mar-09 (F)}
    \label{fig:chi2map_F}
\end{subfigure}
\\
\begin{subfigure}{0.32\textwidth}
        \includegraphics[width=\linewidth]{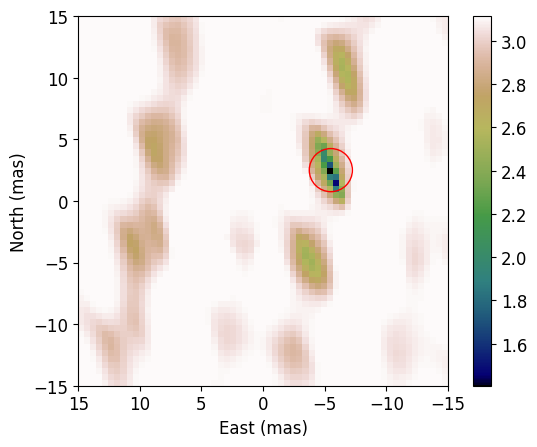}
    \caption{2019-Jun-20 (G)}
    \label{fig:chi2map_G}
\end{subfigure}
\hspace*{\fill}
\begin{subfigure}{0.32\textwidth}
        \includegraphics[width=\linewidth]{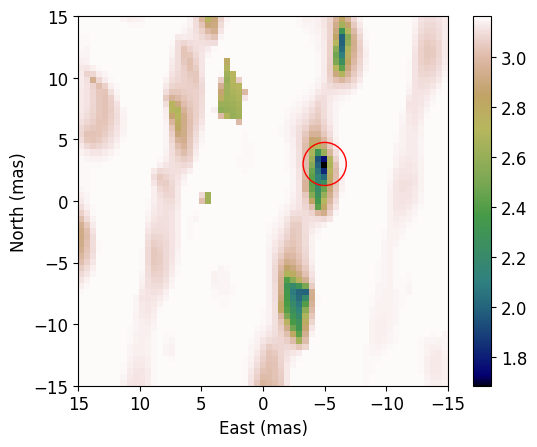}
    \caption{2020-Jan-28 (H)}
    \label{fig:chi2map_H}
\end{subfigure}
\hspace*{\fill}
\begin{subfigure}{0.32\textwidth}
        \includegraphics[width=\linewidth]{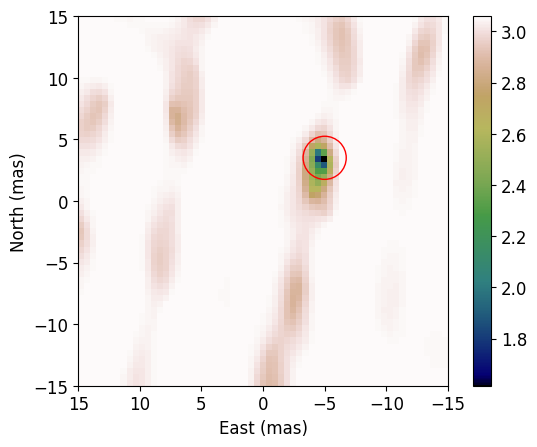}
    \caption{2020-Jan-30 (I)}
    \label{fig:chi2map_I}
\end{subfigure}
\caption{Results of the grid search for the binary position of each epoch. The logarithmic $\chi^2$-maps show a cut through the ($\alpha$, $\beta$)-plane, with the best value of flux ratios at each position. The red ellipses show the position of the global minimum on the map, which coincides with the position of the secondary component selected to derive the orbital parameters. The exception is Run E, where the selected position of the secondary (black ellipse) is not the global minimum but a nearby local minimum.}
\label{fig:chi2maps_grid}
\end{figure*}

%
%
\begin{figure*}
\begin{subfigure}{0.32\textwidth}
        \includegraphics[width=\linewidth]{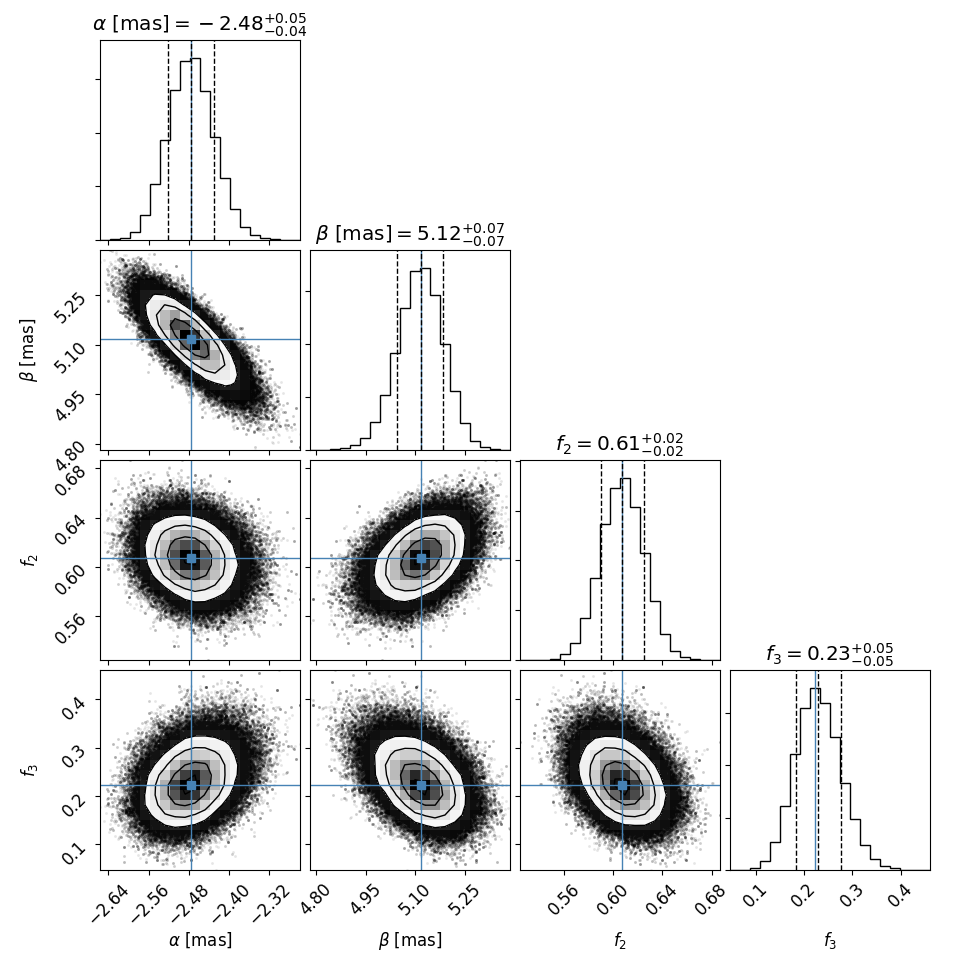}
    \caption{2011-Feb-10 (A)}
    \label{fig:corner_A}
\end{subfigure}
\hspace*{\fill}
\begin{subfigure}{0.32\textwidth}
        \includegraphics[width=\linewidth]{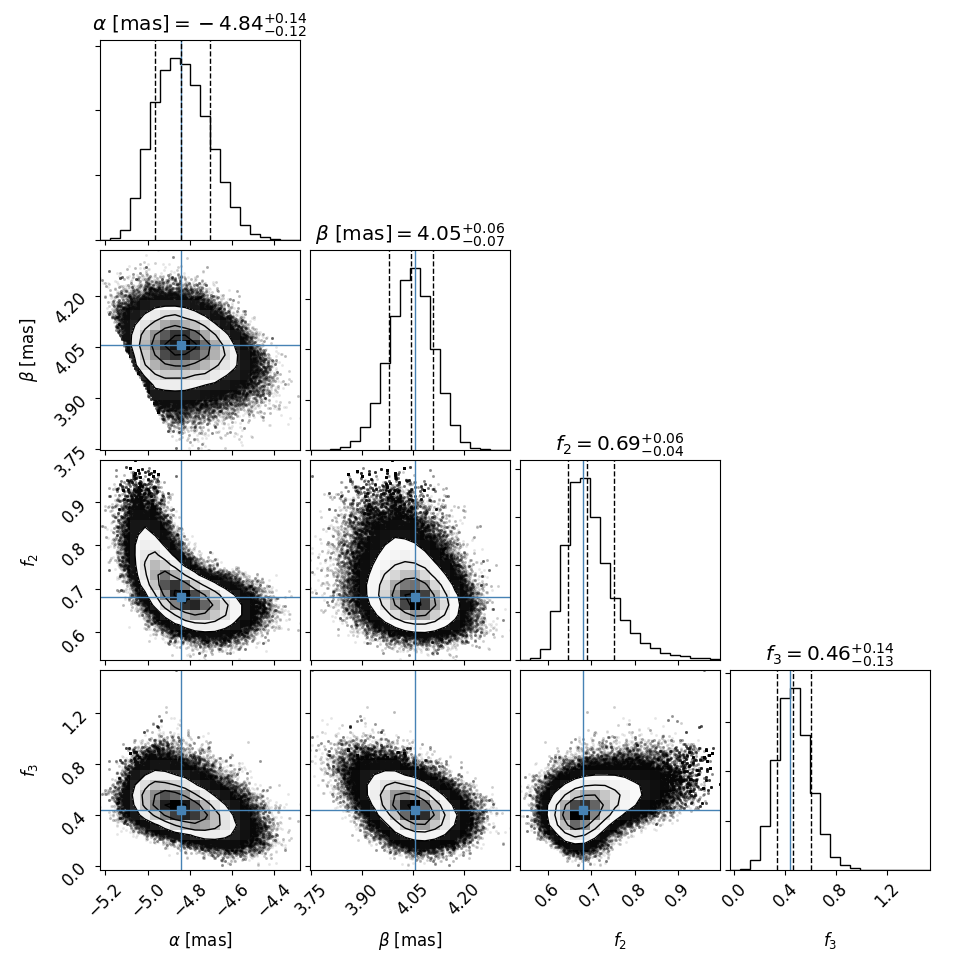}
    \caption{2012-Mar-06 (B)}
    \label{fig:corner_B}
\end{subfigure}
\hspace*{\fill}
\begin{subfigure}{0.32\textwidth}
        \includegraphics[width=\linewidth]{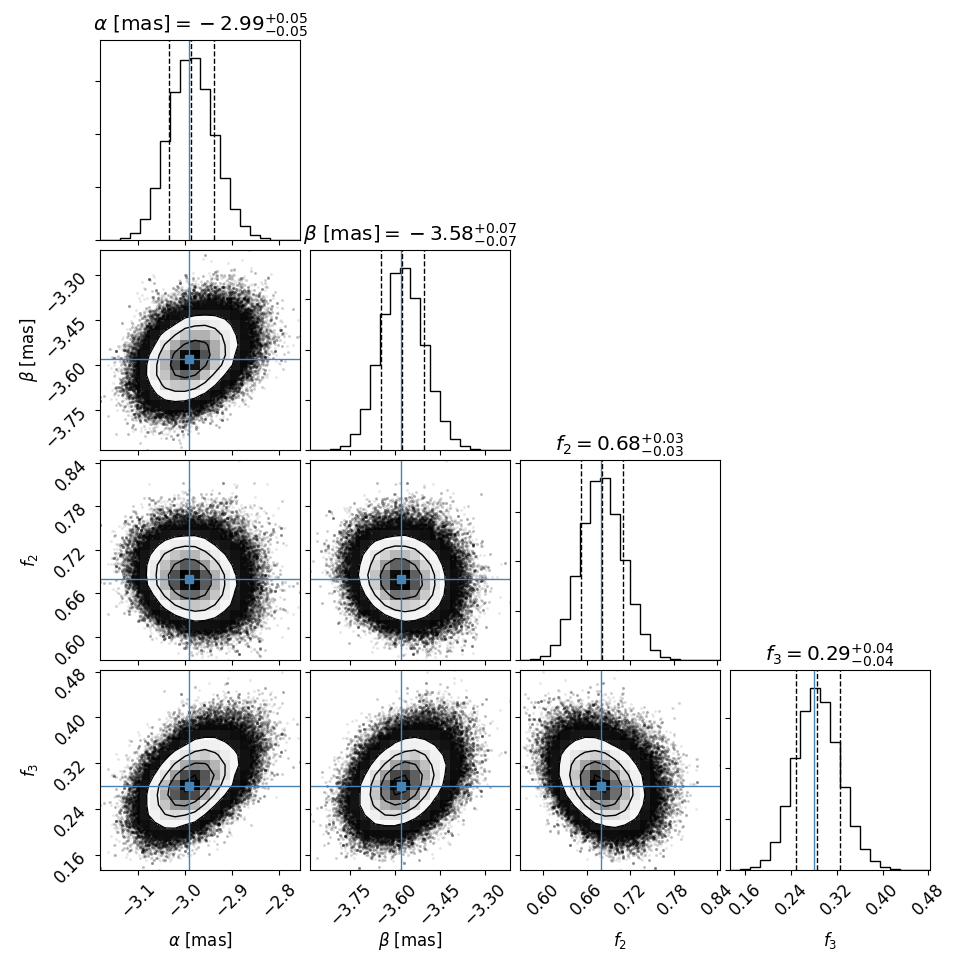}
    \caption{2012-Jul-02 (C)}
    \label{fig:corner_C}
\end{subfigure}
\\
\begin{subfigure}{0.32\textwidth}
        \includegraphics[width=\linewidth]{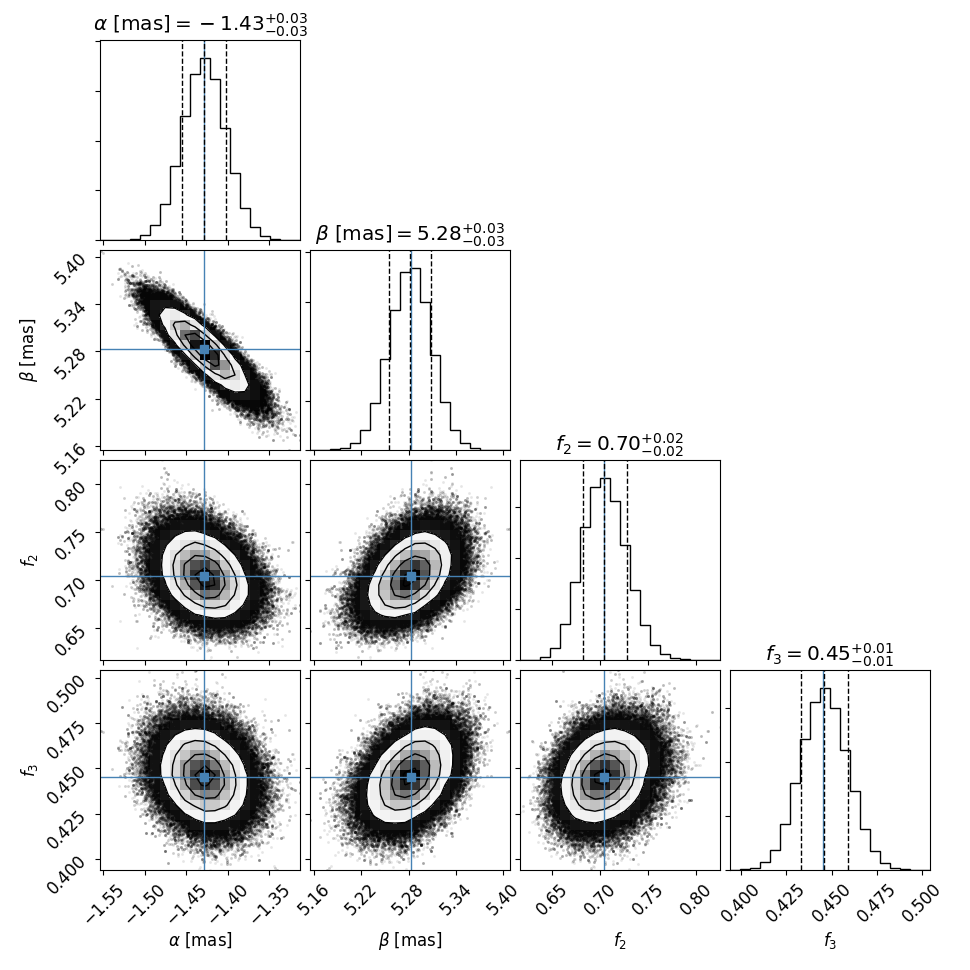}
    \caption{2015-Feb-05 (D)}
    \label{fig:corner_D}
\end{subfigure}
\hspace*{\fill}
\begin{subfigure}{0.32\textwidth}
        \includegraphics[width=\linewidth]{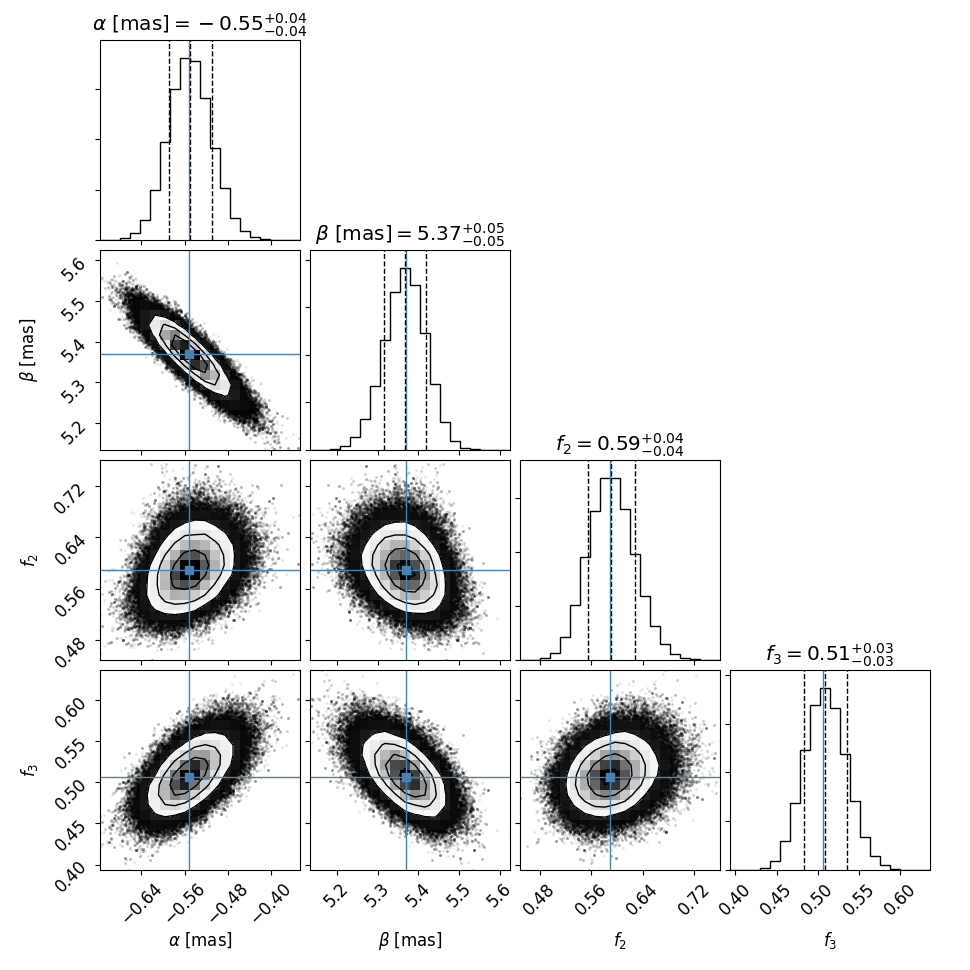}
    \caption{2015-Feb-13 (E)}
    \label{fig:corner_E}
\end{subfigure}
\hspace*{\fill}
\begin{subfigure}{0.32\textwidth}
        \includegraphics[width=\linewidth]{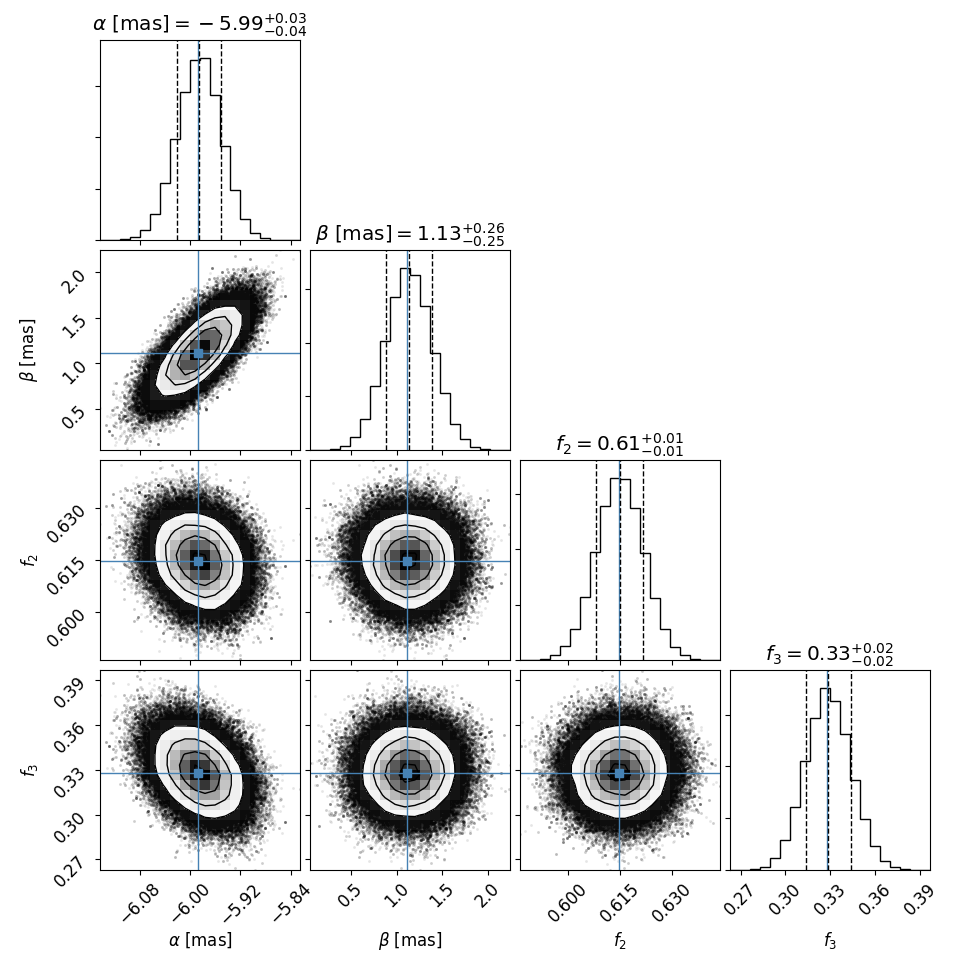}
    \caption{2017-Mar-09 (F)}
    \label{fig:corner_F}
\end{subfigure}
\\
\begin{subfigure}{0.32\textwidth}
        \includegraphics[width=\linewidth]{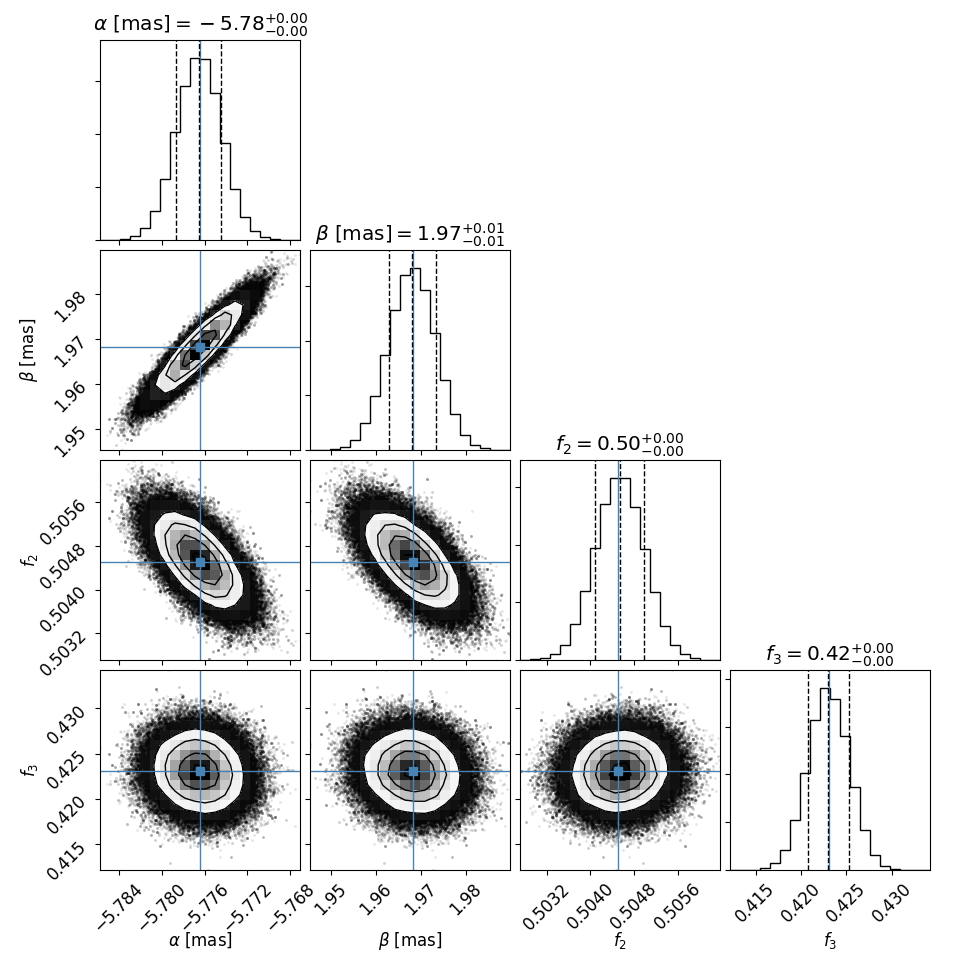}
    \caption{2019-Jun-20 (G)}
    \label{fig:corner_G}
\end{subfigure}
\hspace*{\fill}
\begin{subfigure}{0.32\textwidth}
        \includegraphics[width=\linewidth]{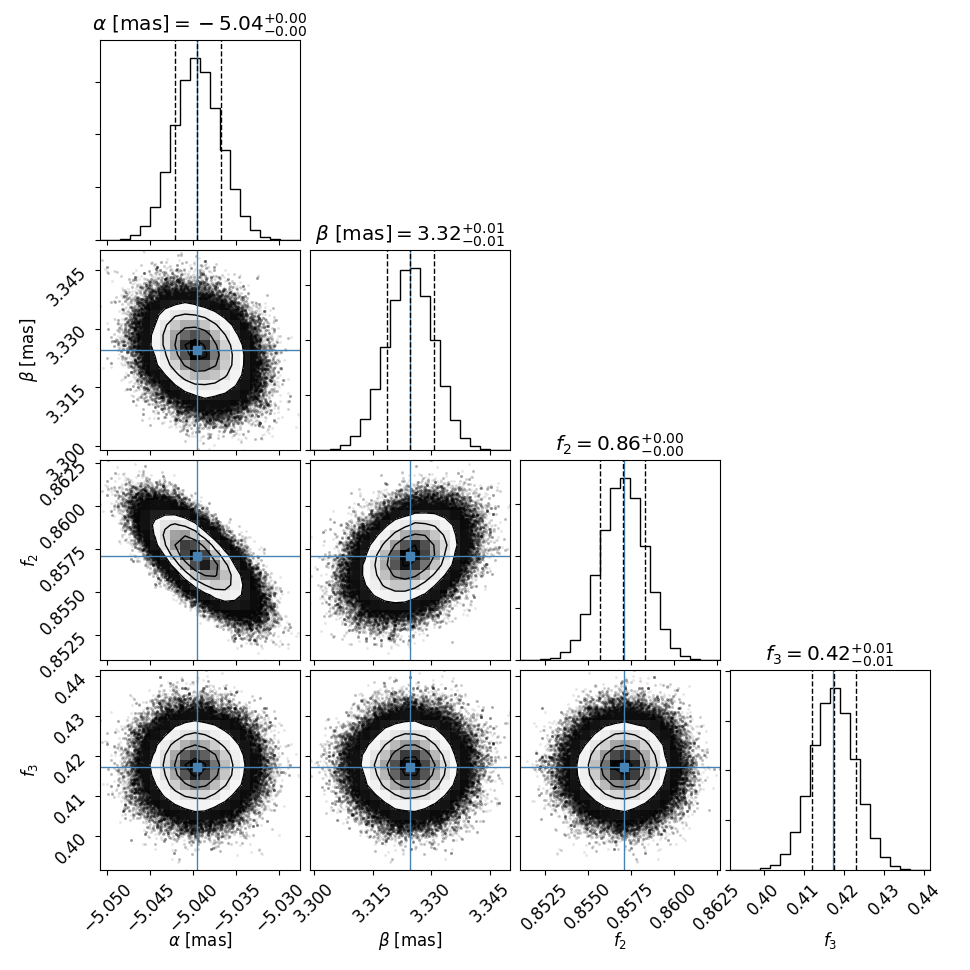}
    \caption{2020-Jan-28 (H)}
    \label{fig:corner_H}
\end{subfigure}
\hspace*{\fill}
\begin{subfigure}{0.32\textwidth}
        \includegraphics[width=\linewidth]{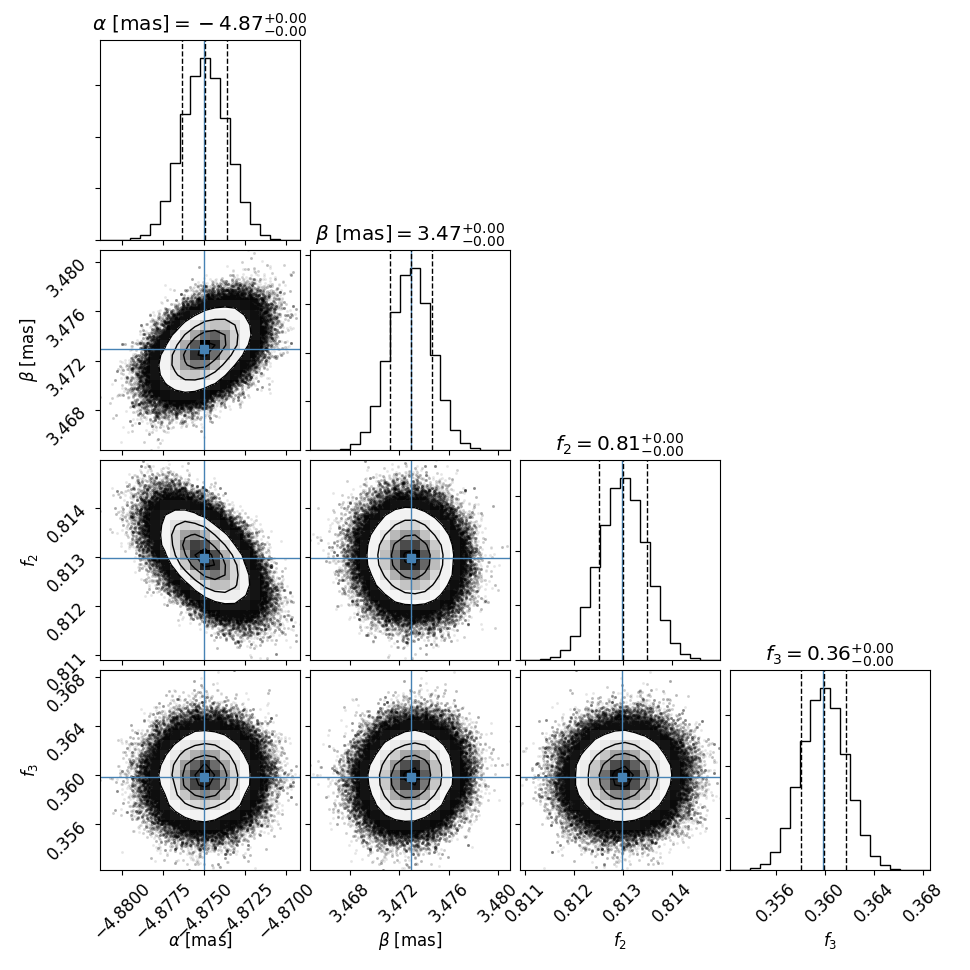}
    \caption{2020-Jan-30 (I)}
    \label{fig:corner_I}
\end{subfigure}
\caption{
        MCMC marginal posterior distributions for the astrometric fit with the four free parameters $\alpha$, $\beta$, $f_2$, and $f_3$, as described in Sect.~\ref{sec:positions}, for each of the nine epochs.
}
\label{fig:positions_MCMC}
\end{figure*}
\clearpage
\newpage
\section{Comparison between raw and binned data from the SC channel of GRAVITY}\label{sec:binned}

\subsection{GRAVITY 2019-Jun-20}
\begin{figure*}
        \begin{subfigure}{0.45\textwidth}
                \includegraphics[width=\linewidth]{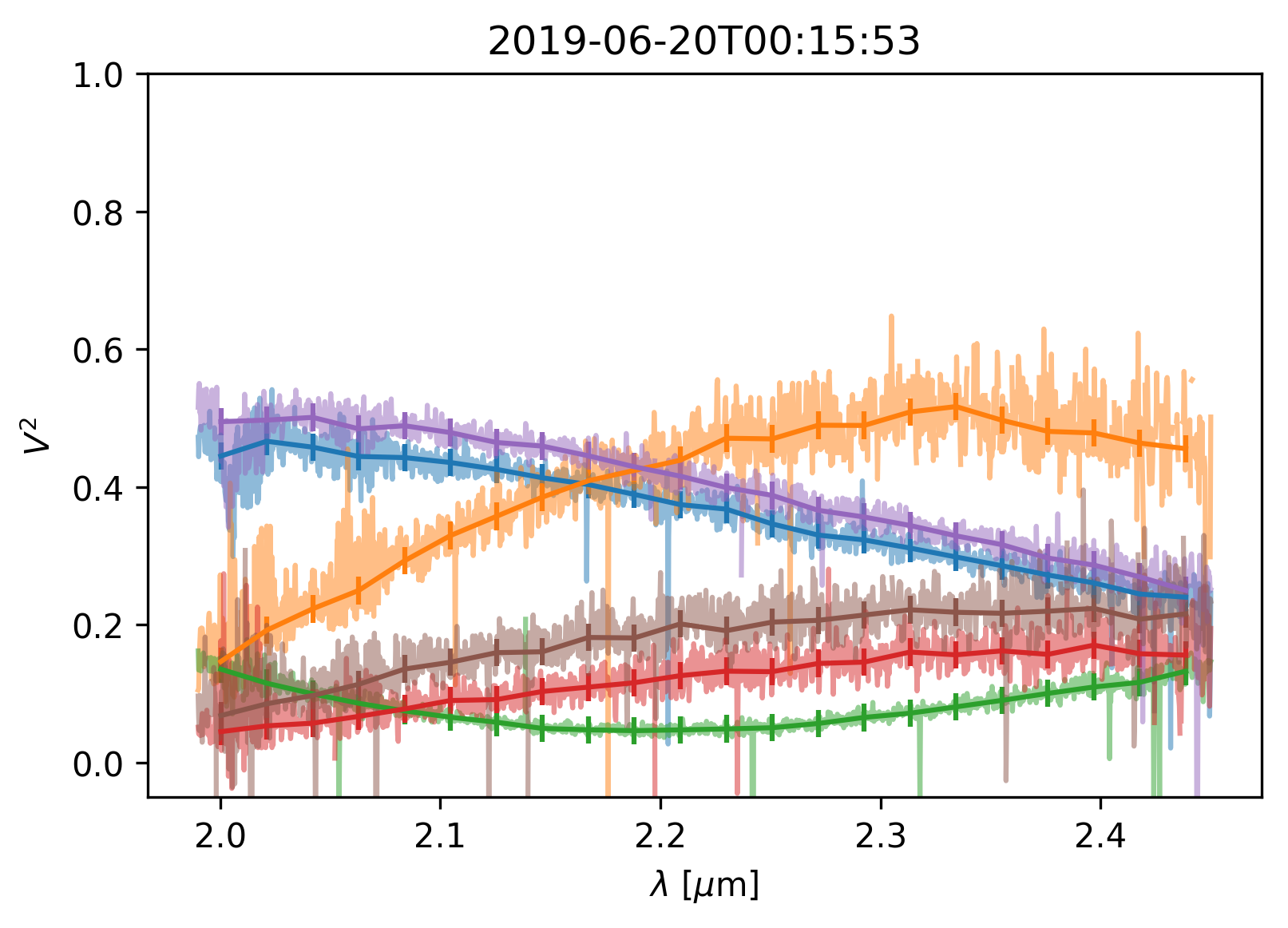}
        \end{subfigure}
        \hspace*{\fill}
        \begin{subfigure}{0.45\textwidth}
                \includegraphics[width=\linewidth]{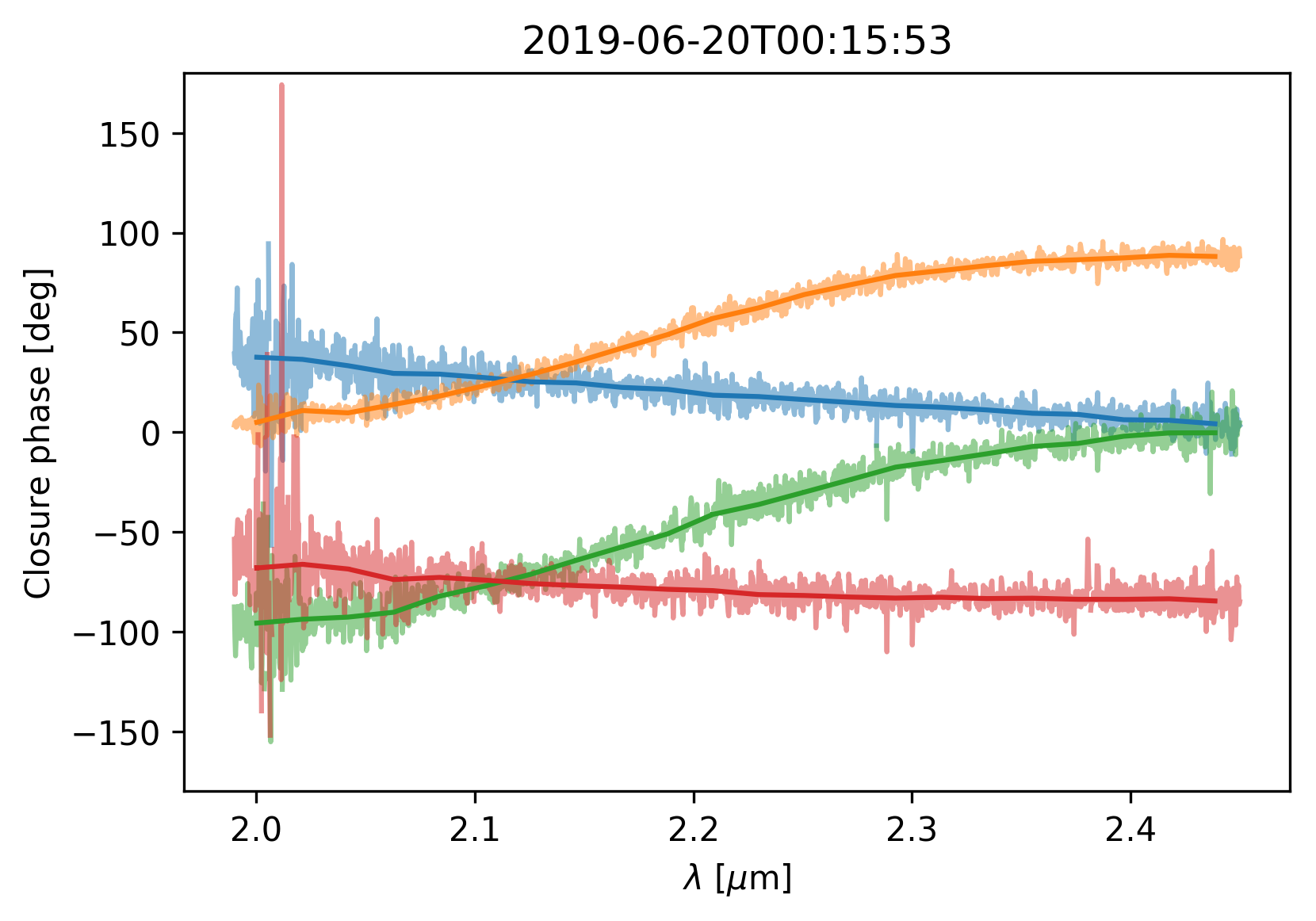}
        \end{subfigure}
        \\
        \begin{subfigure}{0.45\textwidth}
                \includegraphics[width=\linewidth]{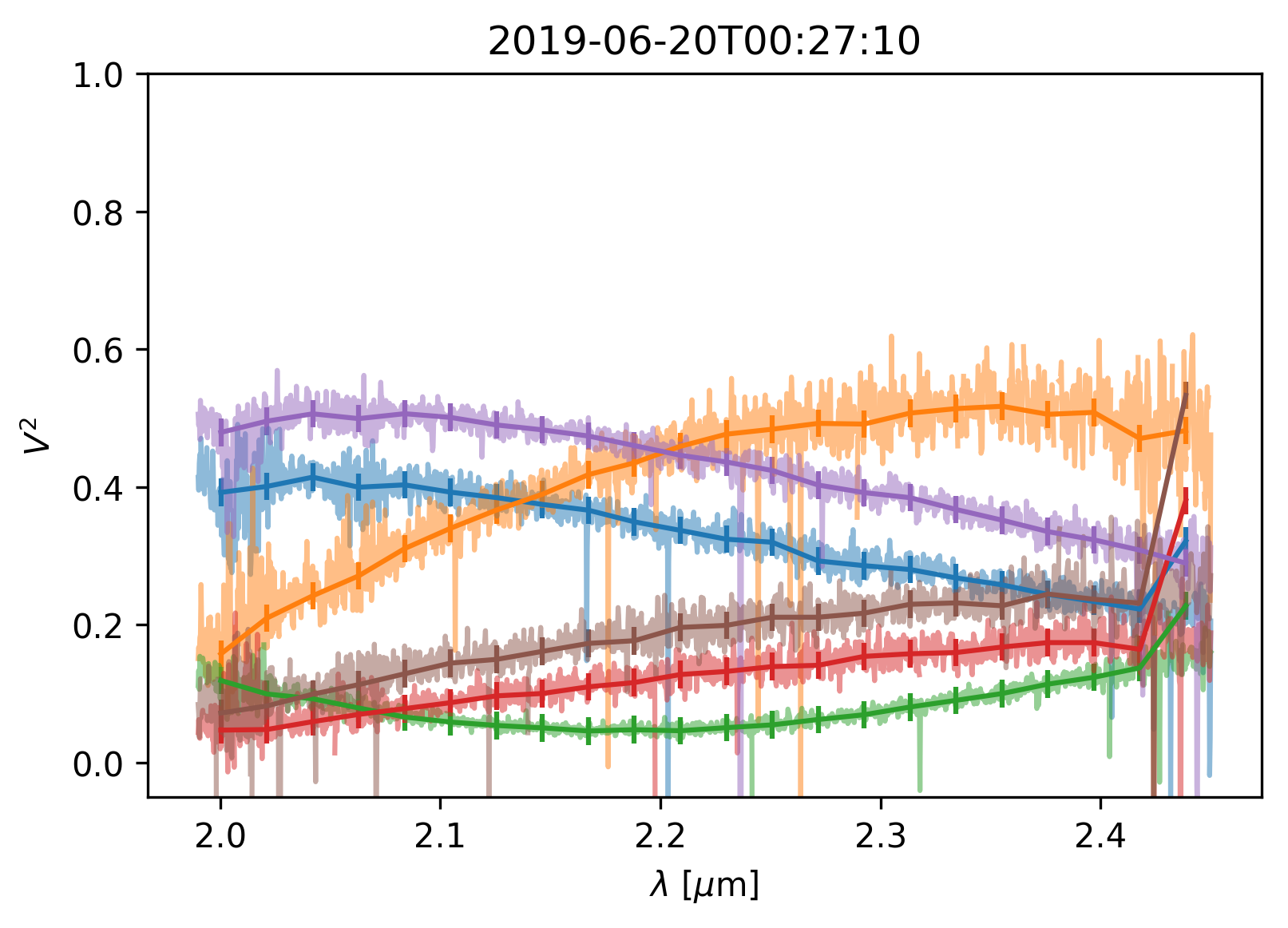}
        \end{subfigure}
        \hspace*{\fill}
        \begin{subfigure}{0.45\textwidth}
                \includegraphics[width=\linewidth]{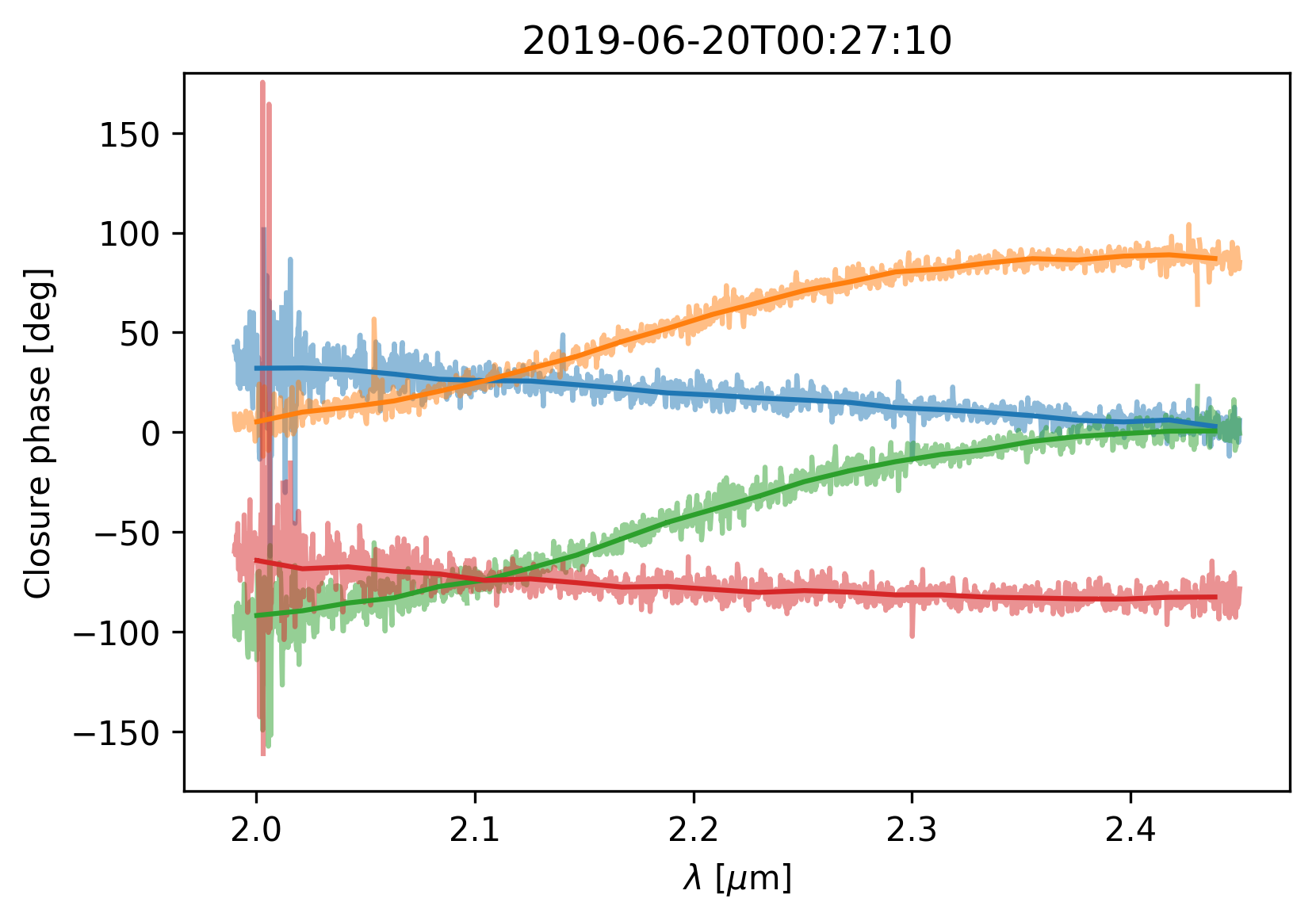}
        \end{subfigure}
        \\
        \begin{subfigure}{0.45\textwidth}
                \includegraphics[width=\linewidth]{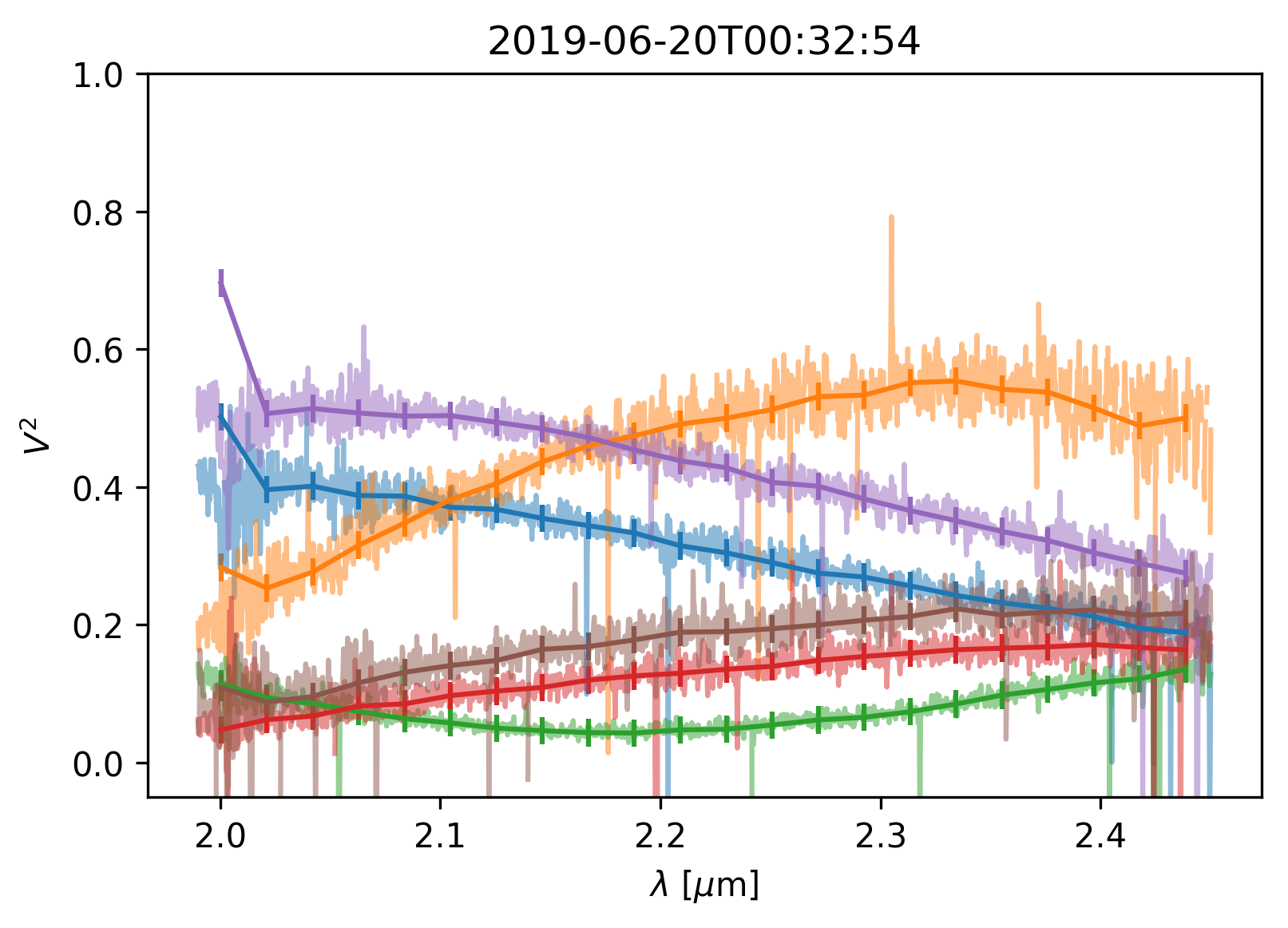}
        \end{subfigure}
        \hspace*{\fill}
        \begin{subfigure}{0.45\textwidth}
                \includegraphics[width=\linewidth]{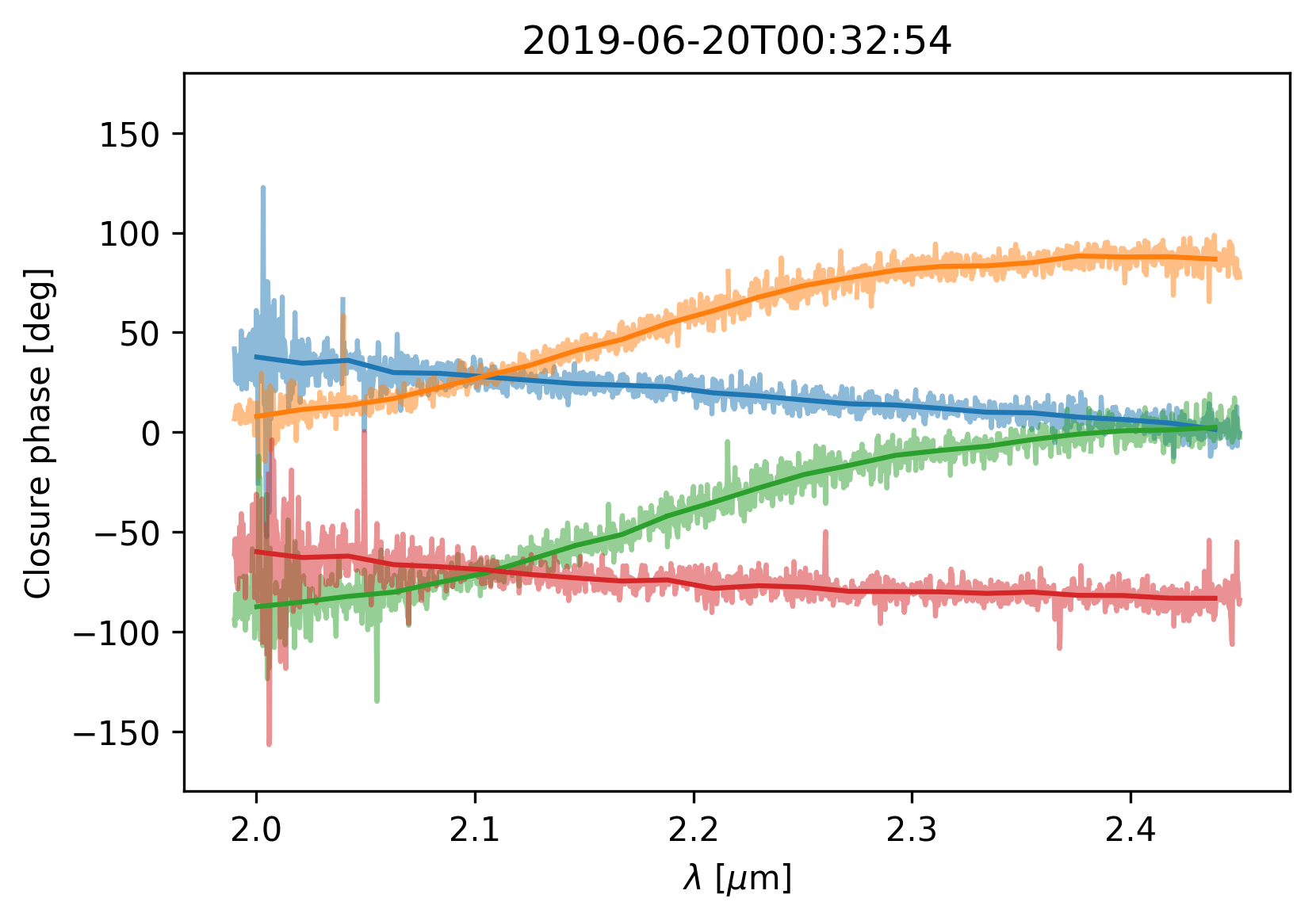}
        \end{subfigure}
        \\
        \begin{subfigure}{0.45\textwidth}
                \includegraphics[width=\linewidth]{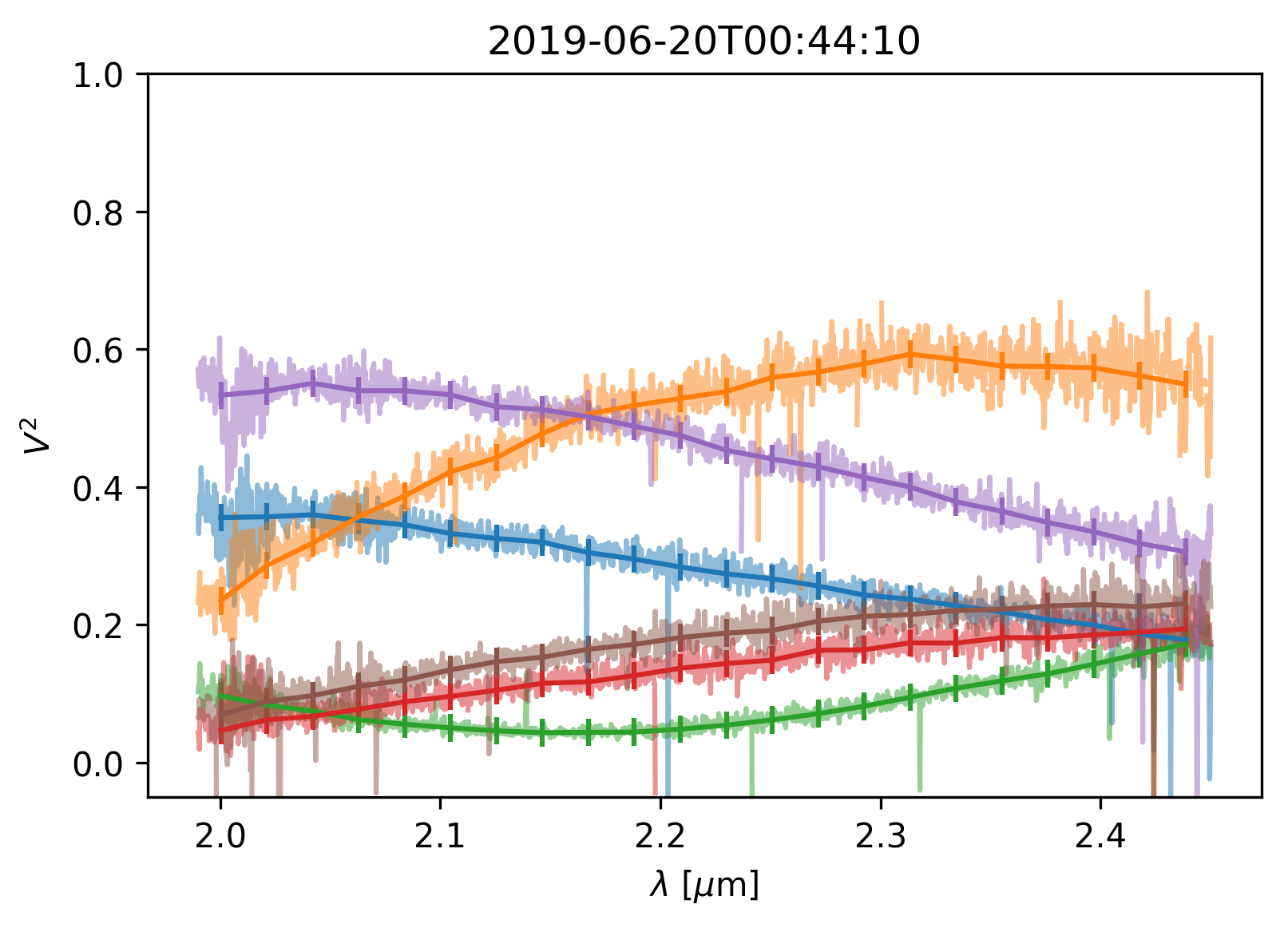}
        \end{subfigure}
        \hspace*{\fill}
        \begin{subfigure}{0.45\textwidth}
                \includegraphics[width=\linewidth]{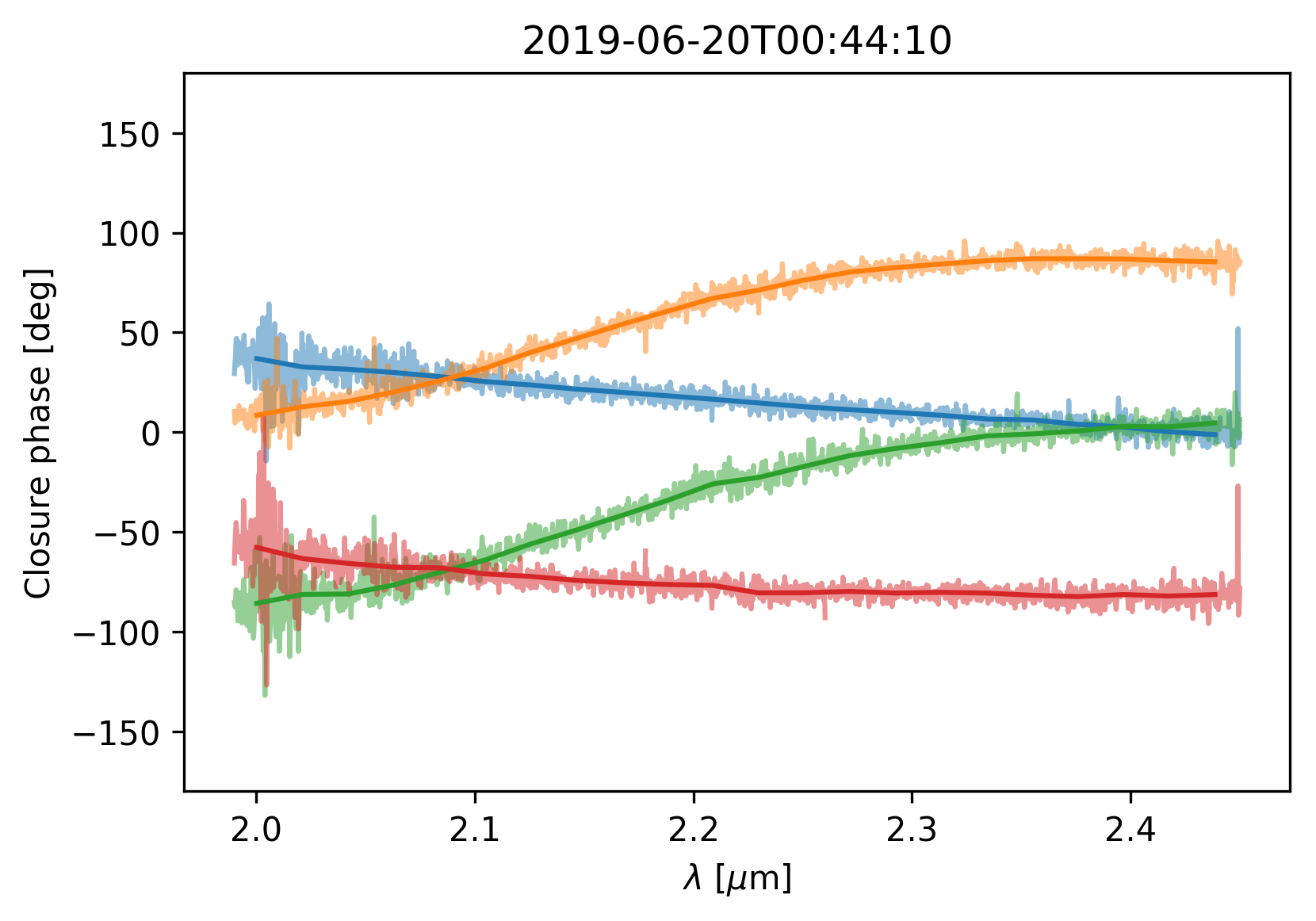}
        \end{subfigure}
        \caption{Comparison of the 2019-Jun-20 squared visibilities and closure phases, as binned by the GRAVITY pipeline (solid lines with error bars) with the full spectral resolution (lighter colors in the background).}
\end{figure*}

\clearpage
\newpage
\subsection{GRAVITY 2020-Jan-30}
\begin{figure*}
        \begin{subfigure}{0.45\textwidth}
                \includegraphics[width=\linewidth]{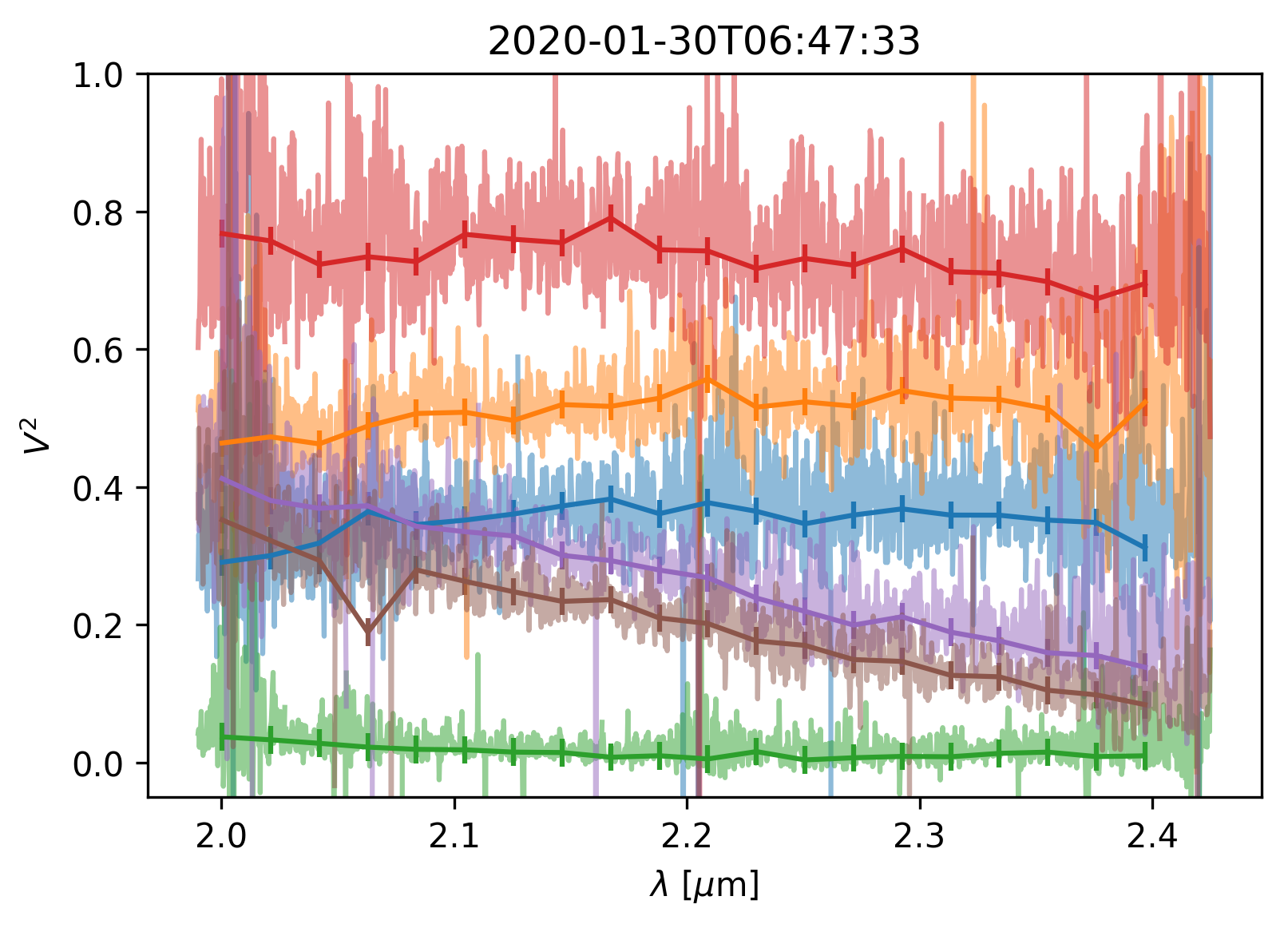}
        \end{subfigure}
        \hspace*{\fill}
        \begin{subfigure}{0.45\textwidth}
                \includegraphics[width=\linewidth]{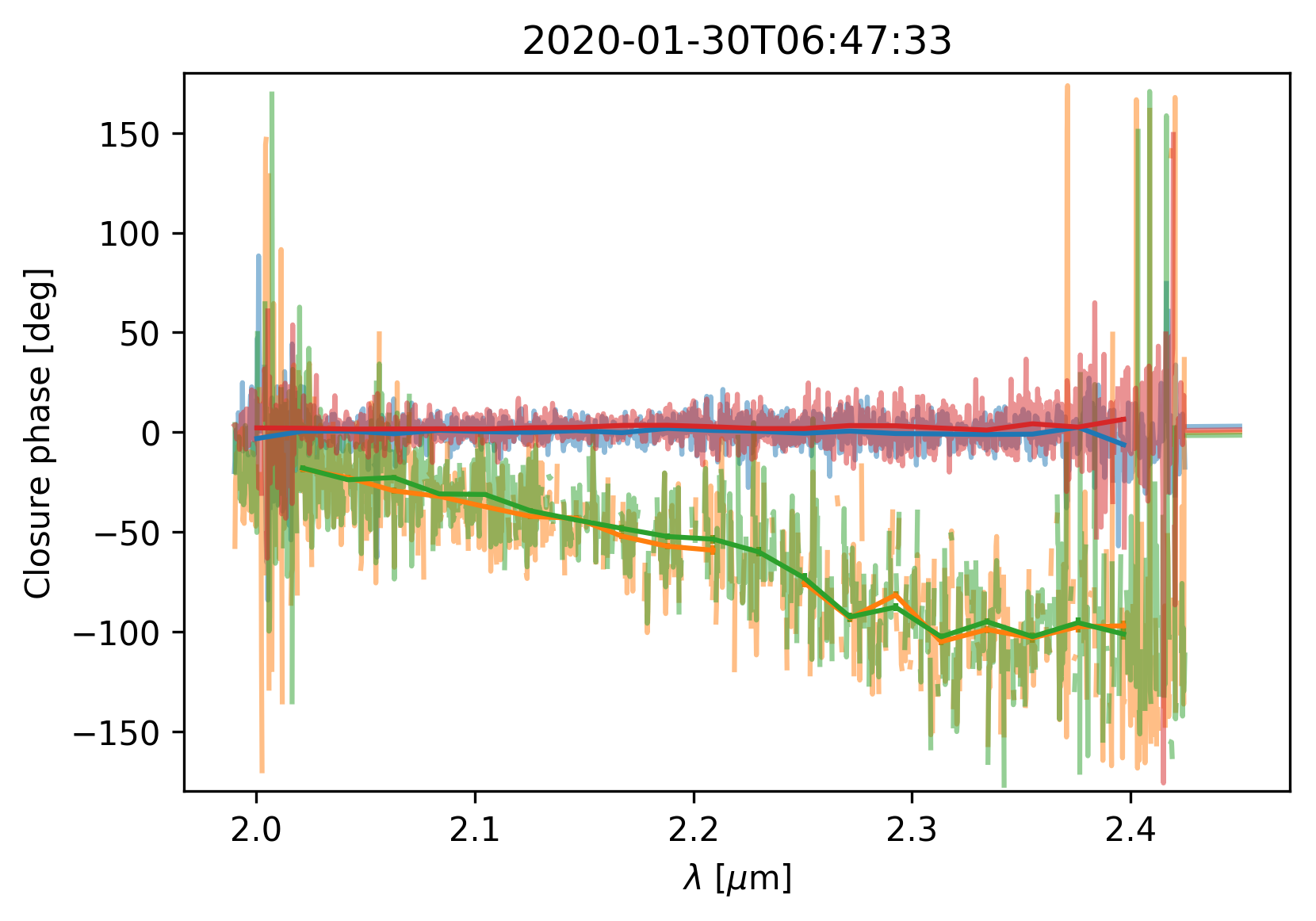}
        \end{subfigure}
        \\
        \begin{subfigure}{0.45\textwidth}
                \includegraphics[width=\linewidth]{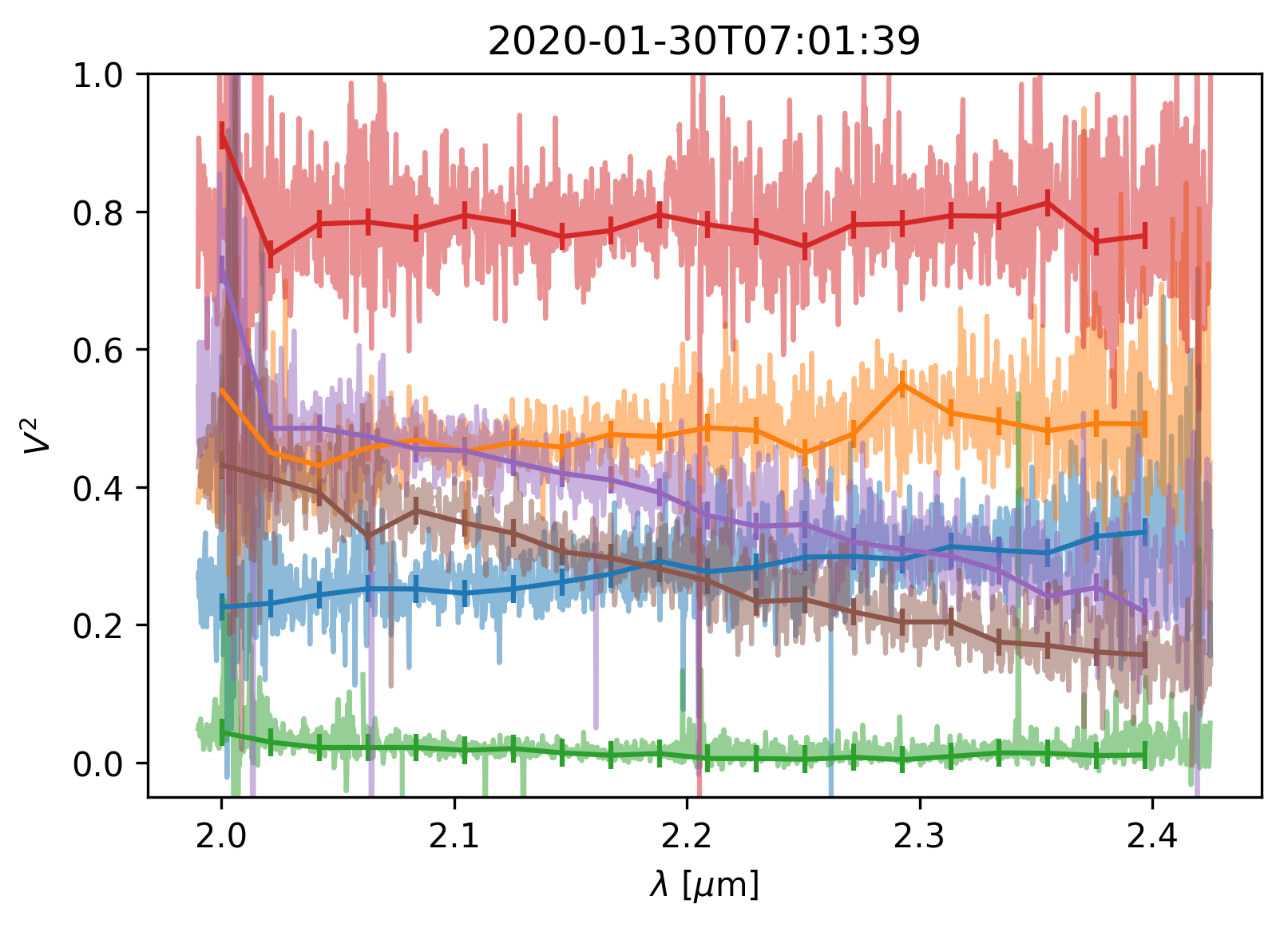}
        \end{subfigure}
        \hspace*{\fill}
        \begin{subfigure}{0.45\textwidth}
                \includegraphics[width=\linewidth]{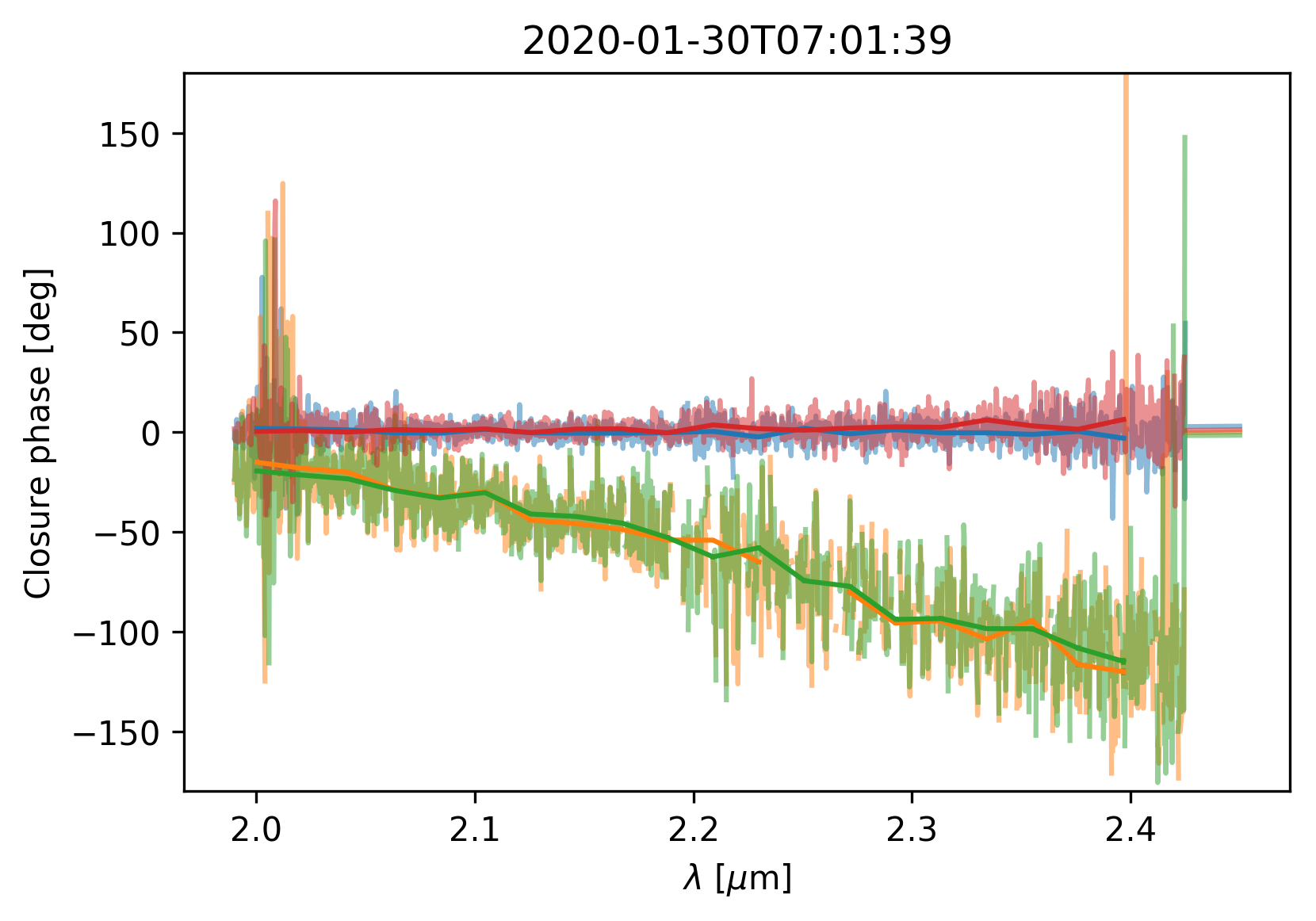}
        \end{subfigure}
        \\
        \begin{subfigure}{0.45\textwidth}
                \includegraphics[width=\linewidth]{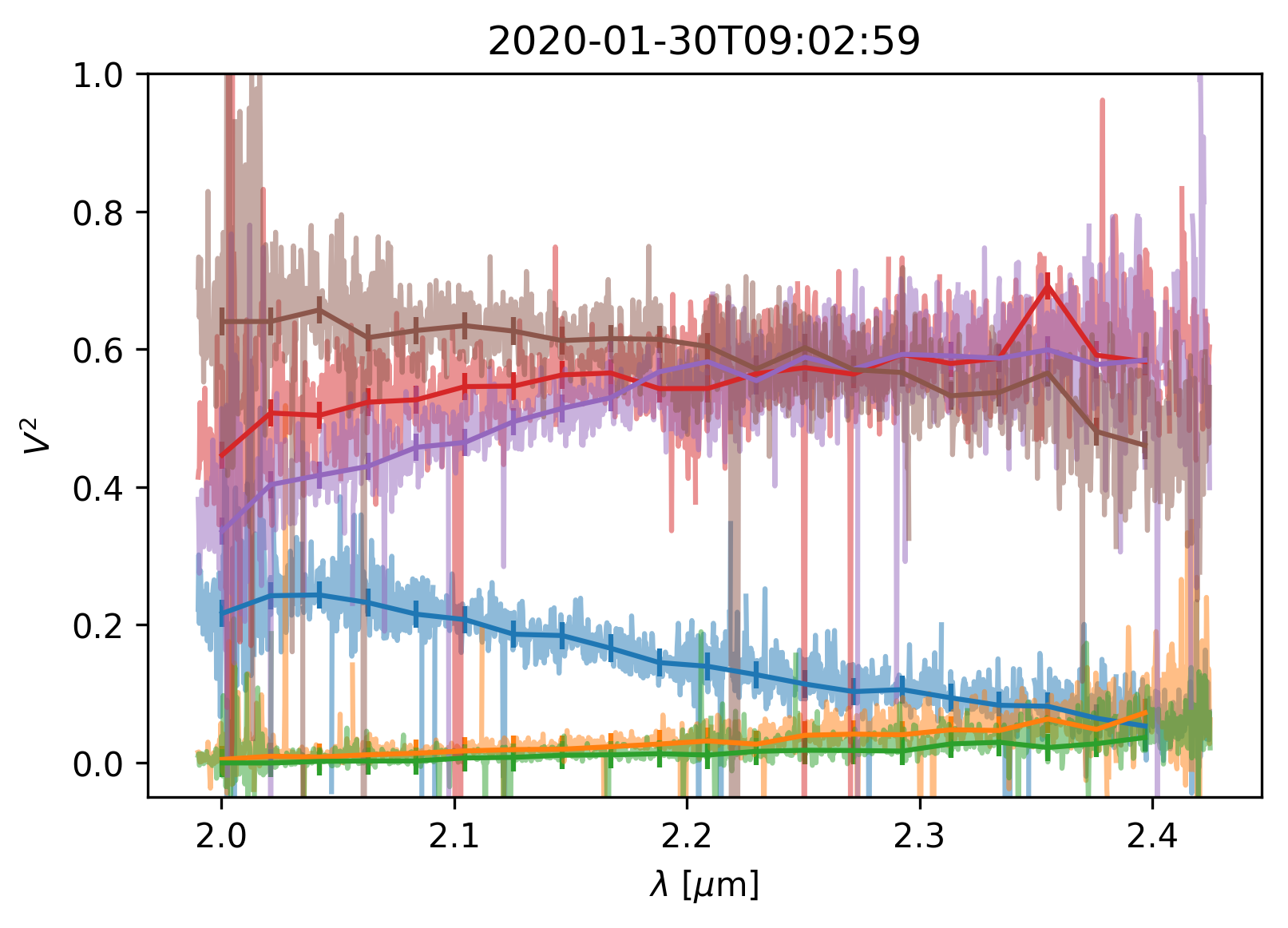}
        \end{subfigure}
        \hspace*{\fill}
        \begin{subfigure}{0.45\textwidth}
                \includegraphics[width=\linewidth]{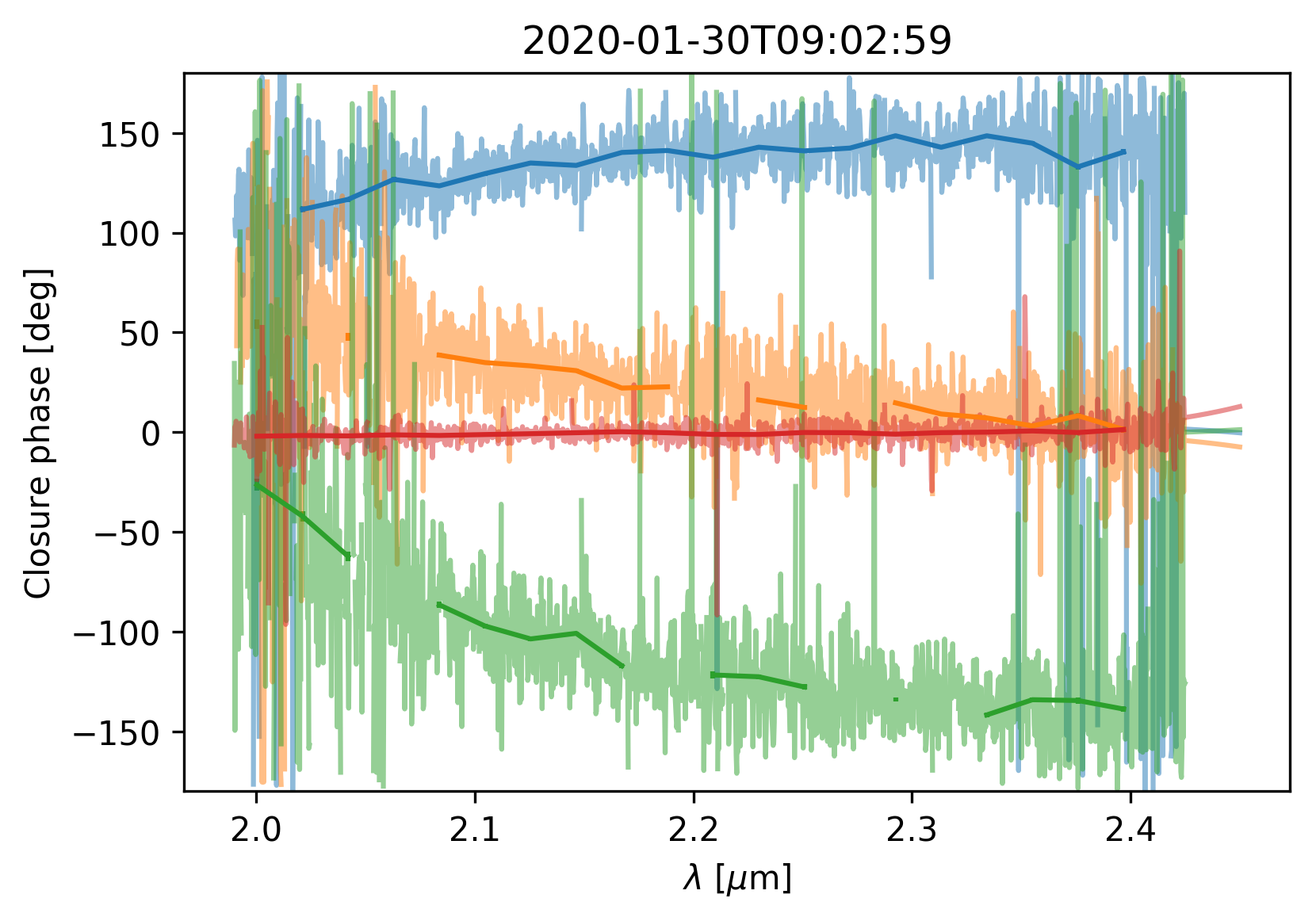}
        \end{subfigure}
        \\
        \begin{subfigure}{0.45\textwidth}
                \includegraphics[width=\linewidth]{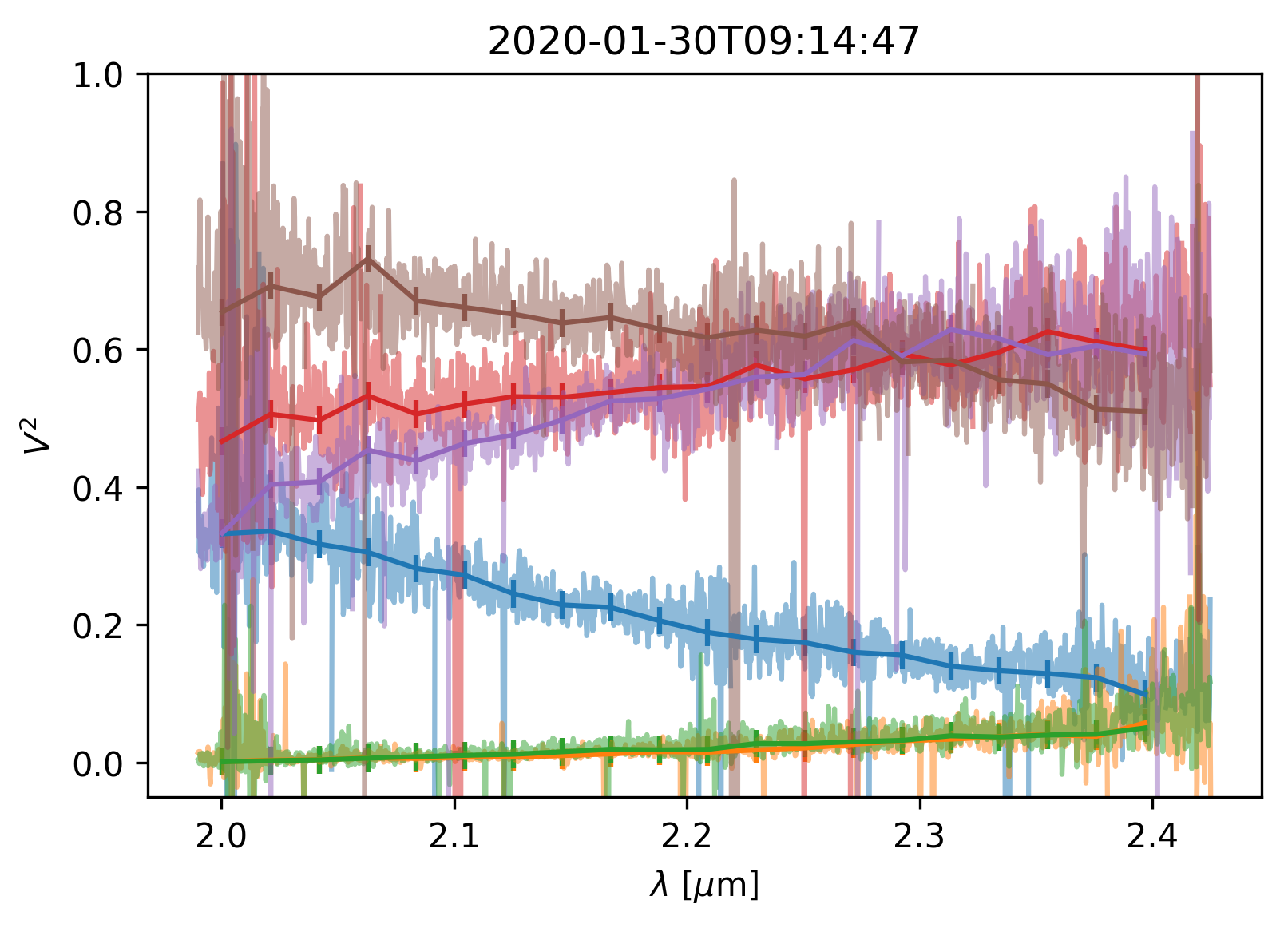}
        \end{subfigure}
        \hspace*{\fill}
        \begin{subfigure}{0.45\textwidth}
                \includegraphics[width=\linewidth]{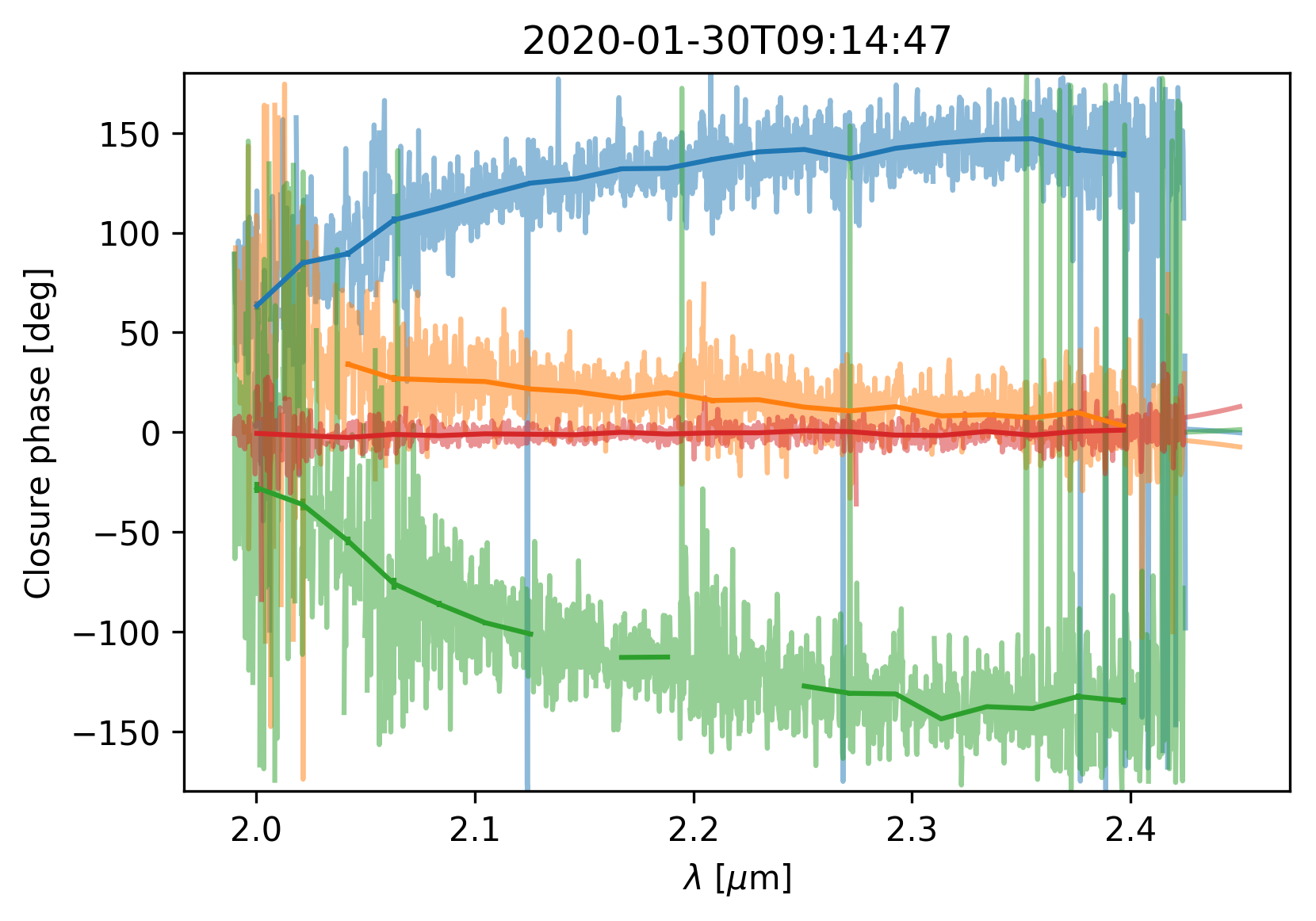}
        \end{subfigure}
        \caption{Comparison of the 2020-Jan-30 squared visibilities and closure phases, as binned by the GRAVITY pipeline (solid lines with error bars) with the full spectral resolution (lighter colors in the background).}
\end{figure*}

\clearpage
\newpage
\section{Parallax and distance of WW\,Cha}\label{sec:parallax}

The precise knowledge of the distance to the source is important in our analysis since it impacts the uncertainty of the derived parameters, such as the dynamical mass. 
With a G magnitude of $\sim$12, the GAIA parallax of WW\,Cha is reported to be $\varpi$\,=\,5.206$\pm$0.035\,mas. Because of the small relative error, $\sigma_{\varpi}$/$\varpi$\,$\sim$\,0.7\%, the naive inverse parallax to estimate the distance is a valid approach (\citet{2018AJ....156...58B}, cf. their Fig.\,6). However, the astrometric solution of Gaia DR2 treats all sources -- including binary stars -- as single stars. \cite{GAIA2018} suggest that for binary stars with periods on the order of 2\,years, the mean Gaia parallax and the true value could differ. We have therefore devoted attention to this aspect.

\begin{figure}[h]
   \centering
   \includegraphics[width=1.0\columnwidth]{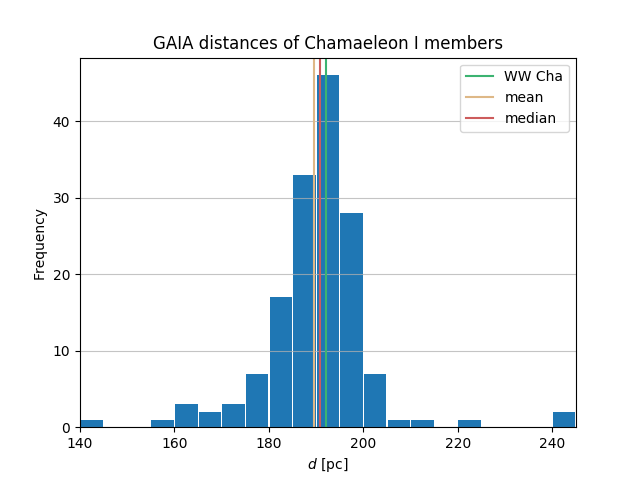}
   \caption{Histogram of Gaia distances for known members of Chamaeleon\,I from \cite{2008ApJ...675.1375L}.}
   \label{fig:parallax}%
\end{figure}

The RUWE\,\footnote{Renormalized unit weight error; \url{https://gea.esac.esa.int/archive/documentation/GDR2/Gaia_archive/chap_datamodel/sec_dm_main_tables/ssec_dm_ruwe.html}} parameter of 1.45 associated with WW\,Cha might be a further indication of the non-single nature of the source with respect to the astrometric solution. This is, however, very close to the proposed limit of 1.4 \citep{Lindegren2018}, below which the measurement can be considered “well behaved” in the DR2 catalog. The shorter $\sim$0.6\,yr period of WW\,Cha also reinforces the reliability of the DR2 record. Following \citet[][cf. their Fig.\,9]{2018AJ....156...58B}, we built the distance histogram (here, the inverse parallax) of identified members of the Chamaeleon\,I star forming region. The plot of Fig.~\ref{fig:parallax} shows the result for a sample of 173 known members of Chamaeleon~I from \cite{2008ApJ...675.1375L} with counterparts in the Gaia DR2 catalog with a 5\,pc bin size. After removing sources below 150\,pc and above 210\,pc, the median of 190.7\,pc and mean of 189.5\,pc (with $\sigma=8.4$\,pc) of the remaining 148 sources are in close agreement with the inverted parallax of WW\,Cha, namely $1/\varpi$\,=\,192.1\,pc. We thus trusted and adapted the distance to WW\,Cha found by \citet{2018AJ....156...58B} and furthermore adapted a distance of $\sim$190\,pc to the Chamaeleon\,I cloud. The latter is only used in this study to reestimate the age of Chamaeleon\,I in Sect.~\ref{sec:intro}. Similarly, the newly adapted distance of WW\,Cha leads to a correction of previously derived luminosities by a factor of $(191\,\mathrm{pc})^2/(160\,\mathrm{pc})^2=1.4$.

Inverting the parallax from Gaia EDR3 (which was published during the reviewing process of this paper) results in a distance to WW\,Cha of $188.8^{+1.1}_{-1.1}$\,pc. This is comparable to the solution obtained by \citet{2018AJ....156...58B} of $191.0^{+1.3}_{-1.3}$\,pc, which we continued to use throughout this paper. Within the error bars, this leaves, for instance, the separation $a$\,[au] unaffected.

\clearpage
\newpage

\section{Image reconstruction of WW\,Cha}\label{sec:img-rec-app}
\begin{figure*}[h]
   \centering
   \includegraphics[width=1.0\textwidth]{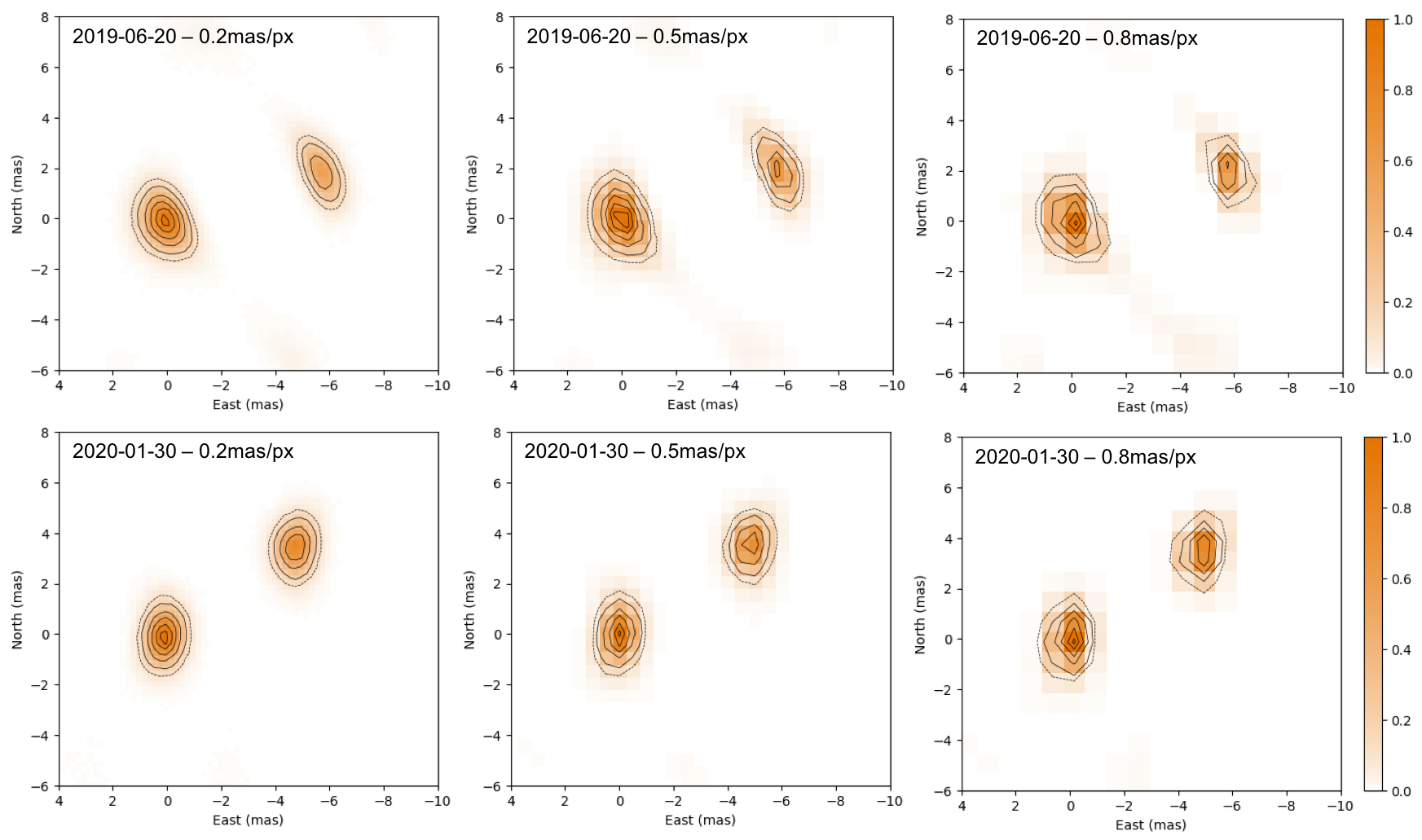}
   \caption{Results of testing the image reconstruction of WW\,Cha using different values for the pixel scale, as indicated, and considering a field of view of 30\,mas.}
   \label{fig:img-rec-app}%
\end{figure*}
\end{document}